\documentclass[prb,aps,twocolumn,floatfix,showpacs ]{revtex4-1}
\usepackage{graphicx}
\usepackage{epstopdf}

\usepackage{color}

\usepackage[utf8]{inputenc}

\begin{document}
\title{Field induced suppression of charge density wave in GdNiC$_2$}

\author{Kamil K. Kolincio, Karolina G\'{o}rnicka, Micha\l{} J. Winiarski, Judyta Strychalska - Nowak, Tomasz Klimczuk} 

\affiliation{Faculty of Applied Physics and Mathematics, Gdansk University of Technology,
Narutowicza 11/12, 80-233 Gdansk, Poland}

\begin{abstract}
We report the specific heat, magnetic, magnetotransport and galvanomagnetic properties of polycrystalline GdNiC$_2$. In the intermediate temperature region above $T_N$ = 20 K, we observe large negative magnetoresistance due to Zeeman splitting of the electronic bands and partial destruction of a charge density wave ground state. Our magnetoresistance and Hall measurements show that at low temperatures a magnetic field induced transformation from antiferromagnetic order to a metamagnetic phase results in the partial suppression of the CDW.
\end{abstract}

\maketitle

\section{Introduction}
The interest in quasi-low dimensional materials lies in their unconventional physical properties. Low dimensionality often results in  anisotropy of thermoelectric and transport properties or electronic instabilities such as charge or spin density waves (CDW and SDW, respectively)
\cite{gruner_density_1994, Monceau2012, Gruner1988}.
The coupling between CDW, magnetic field and magnetic order is a long standing area of interest. In particular, the application of external magnetic field leads to a rich variety of phenomena such as: suppression of CDW due to the Zeeman splitting of the electronic bands\cite{tiedje_magnetoresistance_1975}, enhancement of the CDW \cite{Balseiro1985} or field induced CDW condensation \cite{brooks_magnetic_2006, zanchi_phase_1996, graf2004}.
Since the discovery of the coexistence of CDW and antiferromagnetic order in metallic Cr \cite{Young1974, Fawcett1988}, the extensive efforts have been issued for understanding of the coupling between CDW and magnetism, however the number of compounds exhibiting both Peierls instability and magnetic ordering is limited. In fact, the case of chromium still engages great interest of researchers \cite{jacques2014, nicholson2016, jacques2016}.
Recently, much attention has been devoted to the two families of intermetallic materials: M$_5$Ir$_4$Si$_{10}$, where M = (Er, Yb, Dy, Ho, Y and Tm) \cite{Yang1991, Ghosh1993, galli_charge-density-wave_2000, galli_coexistence_2002, Hossain2005, vansmaalen2004} and RNiC$_2$, where R = (Sm, Tb, Nd and Gd)\cite{hanasaki_magnetic_2012, Yamamoto_2013, Shimomura2016}, in which the emergence of CDW and magnetic ordering has been observed. The study of physical properties of these compounds opens a wide road to explore the interplay between both phenomena. 

GdNiC$_2$ belongs to the group of ternary RNiC$_2$ compounds, forming in the orthorhombic CeNiC$_2$-type structure with a space group of \textit{Amm2}
\cite{bodak_o._i.__1979,jeitschko_ternary_1986, matsuo_antiferromagnetism_1996}. In this system, magnetic order originates entirely from the 4$f$ electrons of rare-earth elements, while Ni atoms have been found to carry no magnetic moments
\cite{onodera_magnetic_1998,KOTSANIDIS_1989}. For R = Sm, the magnetic ground state is ferromagnetic (FM), while compounds with R = (Gd, Tb and Nd) show antiferromagnetic (AF) character\cite{onodera_magnetic_1998,ONODERA_1995,Schafer_1997}. 
 The anomalous temperature dependence of electrical resistivity and lattice constants of RNiC$_2$ \cite{murase}  have been identified as genuine Peierls transitions associated with partial nesting of the Fermi surface (FS) built of warped sheets perpendicular to $a$ axis\cite{laverock_electronic_2009, kim_chemical_2013}. This charge density wave instability, associated with distortion of Ni atoms   \cite{wolfel_commensurate_2010},  is  accompanied with opening of an electronic gap and condensation of a certain portion of electronic carriers \cite{ahmad_evidence_2015}.

GdNiC$_2$ shows a CDW transition with Peierls temperature $T_P$ = 205 K. Initially, the FS nesting occurs with slightly incommensurate wavevector $q = \left[ \frac{1}{2}; \eta; 0\right]$, which evolves into a doubly commensurate value of $q =\left[ \frac{1}{2}; \frac{1}{2}; 0\right] $ \cite{Shimomura2016}. A recent X-Ray diffuse scattering study of the satellite reflections \cite{Shimomura2016} has shown that, in contrast to SmNiC$_2$, the CDW in GdNiC$_2$ survives the transition to a magnetically ordered (AF) state at $T_N$ = 20 K. Owing to the rich phase diagram of GdNiC$_2$ \cite{Hanasaki2011}, it becomes interesting to study the CDW response to the application of external magnetic field and the evolution of magnetic order. Here we report the magnetoresistance, Hall effect, magnetization and specific heat measurements and discuss the destructive influence of magnetic field and magnetic transitions on the CDW in GdNiC$_2$.

\section{Experimental}
A polycrystalline sample of GdNiC$_2$ was synthesized by arc-melting stoichiometric amounts of elemental precursors (from Alfa Aesar): gadolinium (99.9\%), nickel (99.999\%) and carbon (99.997\%). Melting took place in a water-cooled copper hearth, under an ultra-high purity argon atmosphere.  A zirconium button was used as an oxygen getter. To homogenize the specimen, the obtained sample was remelted four times. The arc-melted button was wrapped in tantalum foil, placed in an evacuated quartz tube and annealed at 900{$^\circ$}C for 10 days and quickly cooled down by water-ice quenching. 

Magnetization measurements were carried out using the AC Susceptibility Option (ACMS) of a Quantum Design Physical Properties Measurement System (PPMS). A piece of the sample was fixed in a standard polyethylene straw holder.

Thin slices of the sample for transport and Hall effect measurements were cut with wire saw and polished. Platinum wires serving as electrical leads were spark-welded to the sample surface. The experiments were performed using the PPMS. Resistivity was measured employing the standard four probe technique. The Hall voltage was measured in the direction perpendicular to the electrical current in a presence of magnetic field perpendicular to the sample surface. The data were collected reversing the orientation of applied magnetic field, in order to subtract the parasitic longitudinal magnetoresistance voltage component due to small misalignment of contacts.

Specific heat measurements were done by means of the standard $2\tau$ relaxation method of the PPMS system on a flat polished sample (approx. 4.5 mg).
  
\section{Results and discussion}

 \begin{figure*}[ht]
 \includegraphics[scale=0.6]{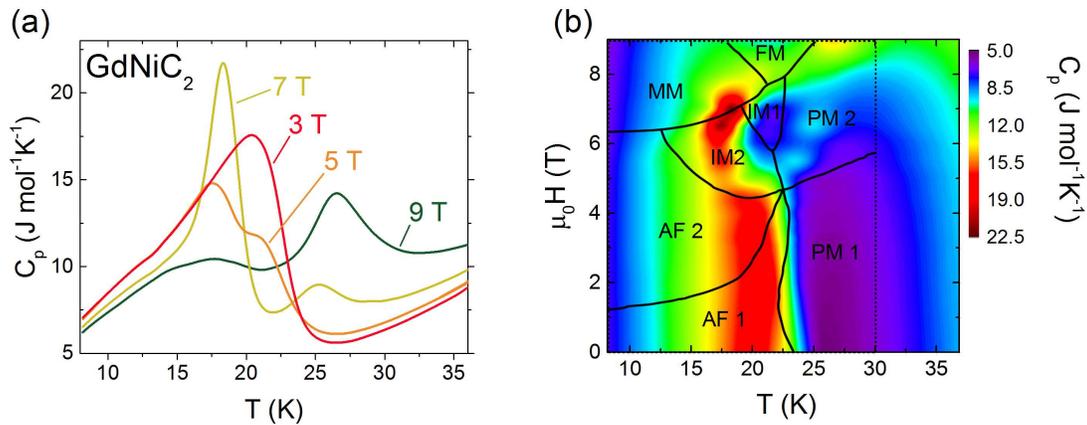}
  \caption{\label{CP} (a) The dependence of specific heat $C_p$ on temperature in applied magnetic field $\mu_0 H$ of 3, 5, 7, and 9 T. 
(b) Map of specific heat of GdNiC$_2$ as a function of temperature and applied magnetic field. $C_p$ vs. $T$ measurements at constant $H$ were used to construct the plot. The phase diagram proposed by Hanasaki et al. \cite{Hanasaki2011} based on magnetization measurements is superimposed on the experimental data (black lines). }
  \end{figure*}

\begin{figure}[ht]
 \includegraphics[scale=0.37]{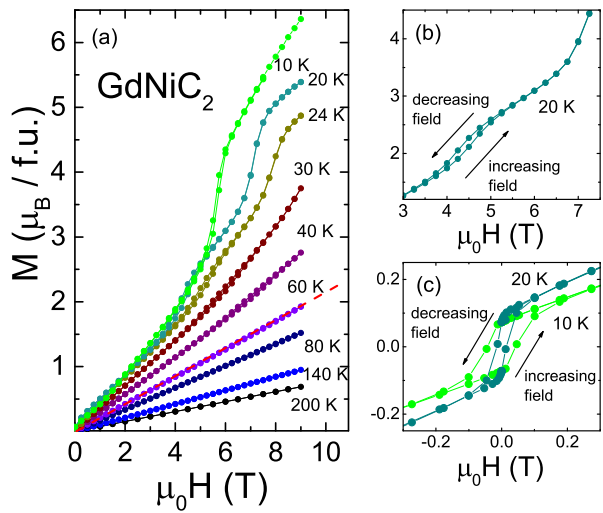}
  \caption{\label{moment}(a) Magnetisation of GdNiC$_2$ measured at various temperatures as a function of $H$. Panels (b) and (c) show the expanded views of hysteresis observed at high and low fields, respectively.}
  \end{figure}
  The purity and crystallographic structure of the sample was tested by powder X-ray diffraction  - see the Supplemental Material. 
  The results of specific heat measurements in the vicinity of the antiferromagnetic transition are shown in fig. \ref{CP} compared with the magnetic phase diagram proposed by Hanasaki et al. \cite{Hanasaki2011} based on magnetizatin measurements. The temperature of a transition from paramagnetic (PM 1) to antiferromagnetic (AF 1) phase is almost unaffected by applied magnetic field up to ca. 4.5 T, above which the specific heat peak at the N\'eel temperature ($T_N$ = 20 - 23 K) is suppressed and two new peaks appear (see Fig. \ref{CP}(a)). Between 5 and 7 T two peaks are seen gradually shifting towards higher temperatures with increasing external magnetic field.
  
While the small change of $T_N$ with applied field is consistent with the phase diagram of Hanasaki et al. \cite{Hanasaki2011}, the transition between the different AF and PM phases (AF 1 - AF 2 and PM 1 - PM 2, see Fig. \ref{CP}) could not be observed within the available measurement accuracy. The phase boundaries between paramagnetic PM 2 and 'intermediate' phases IM 1 and IM 2 seem to be in qualitative agreement with peaks positions shifted towards higher temperatures. A large $C_p$ peak arising at 9 T between 25 and 27 K can likely be attributed to the field-induced ferromagnetic transition observed previously  in magnetization measurements \cite{Hanasaki2011} at slightly lower fields and temperatures. These differences may be caused by the effects of crystal structure disorder that are much larger in a polycrystalline sample than in a single-crystal. 

Detailed analysis of heating-cooling curves recorded by the PPMS calorimeter have not revealed any discernible distortions that should be seen in case of first order phase transitions \cite{PPMS_HC_manual}. No significant peak of specific heat was observed at $T_P$, which is  typical of CDW transition with small lattice deformation.

Results of magnetization measurements vs. applied field ($M$ vs. $\mu _0 H$) are presented in Fig. \ref{moment}. At temperatures down to 60 K the sample magnetization shows a linear behavior without hysteresis. No features are observed in $M$ vs. $T$  (not shown) at $T_P$, which is not surprising, since the change of Pauli - Landau magnetic components are expected to be significantly weaker than the Curie - Weiss term from the local strong magnetic moments. Between 60 K and the N\'{e}el temperature $T_N$, which is slightly above 20 K, curves start to depart from linearity and a small hysteresis loop is formed between approximately 4 and 8 T. Below the $T_N$ two hysteresis loops are observed - larger, between 3 and 6 T (see Fig. \ref{moment}(b)) and smaller at fields up to ca. 0.2 T (see Fig. \ref{moment}(c)). The low-field hysteresis can either be attributed to a previously overlooked phase transition or to a trace amount of ferromagnetic impurity; however lack of an additional specific heat anomaly seems to support the latter. It is interesting that the high-field hysteresis starts to develop already between 40 and 30 K, well above the $T_N$. At 10 K the curve does not saturate even at 9 T, in agreement with previous reports \cite{Onodera_1997}, where the saturation field at 4.2 K was found to be 9.7 or 12 T, depending on the crystallographic axis.

\begin{figure}[h]
 \includegraphics[scale=0.32]{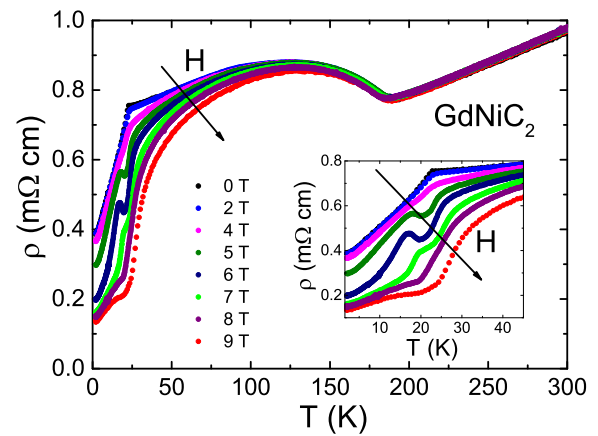}
  \caption{\label{RT} Thermal dependence of resistivity of GdNiC$_2$ at various magnetic fields. Inset: expanded view of the range corresponding to low temperature phase transitions.}
  \end{figure}
  
   \begin{figure}[h]
 \includegraphics[scale=0.32]{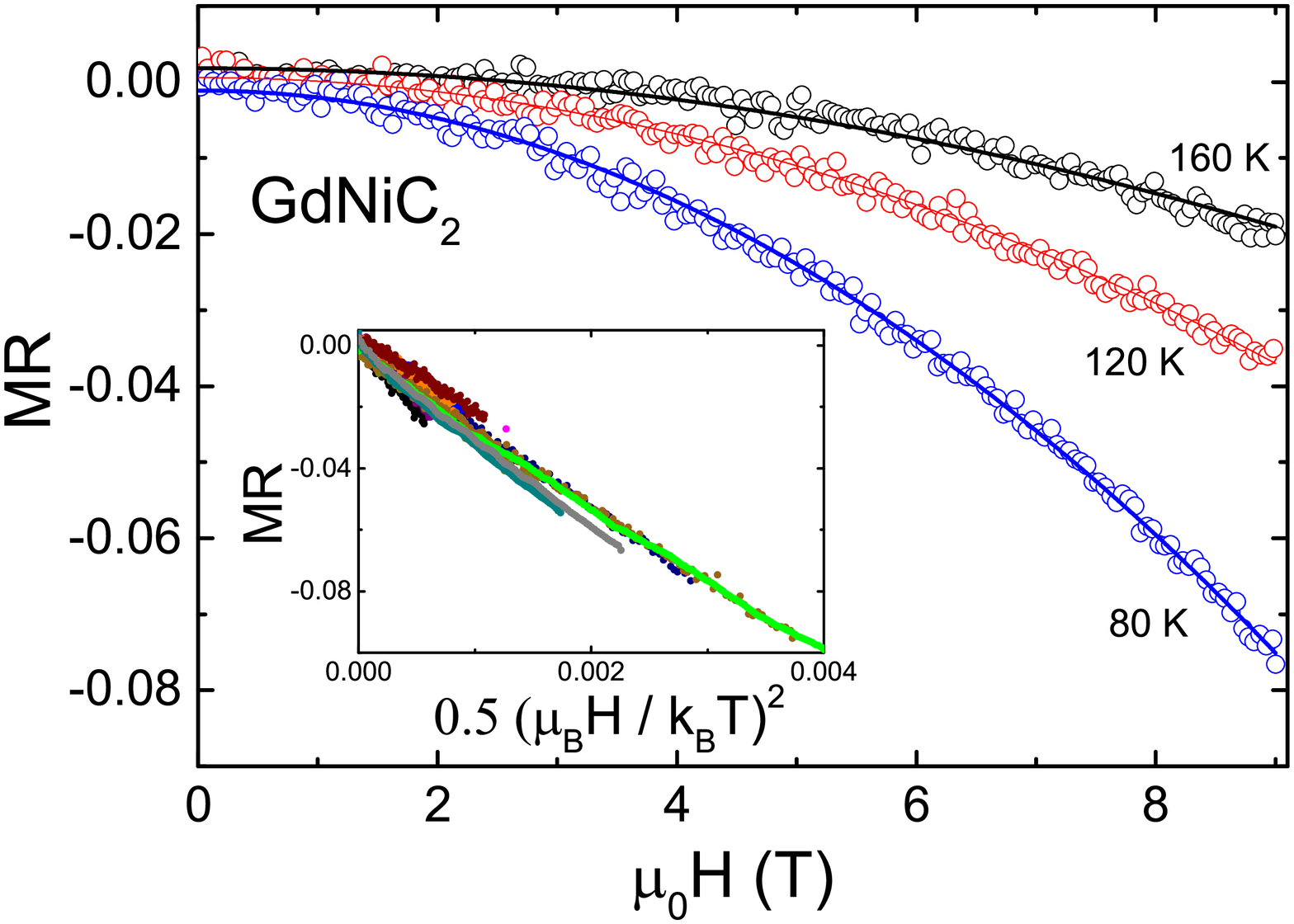}
  \caption{\label{MR} Magnetoresistance of GdNiC$_2$, as a function of magnetic field. Solid lines correspond to $-H^2$ fits to the experimental data. Inset: scaling of MR with eq. \ref{Zeeman} for 30 K $ \leq T \leq $ 180 K.}
  \end{figure}
  
We have used the transport measurements to explore the influence of magnetic field and magnetic transitions to CDW. The main panel of figure \ref{RT} shows the thermal dependence of electrical resistivity of GdNiC$_2$ measured at various magnetic fields. At high temperatures, the zero field resistivity ($\rho_0$) shows typical metallic behavior. At $T_P$ = 196 K, a clear anomaly is pronounced as a metal - metal transition, which is characteristic for the CDW transition in a quasi-2D material, in which the nesting of the Fermi surface is imperfect with fragments of FS still remaining below $T_P$ \cite{kolincio}.  At $T$ = 20 K, which corresponds to $T_N$, one observes an abrupt decrease of resistivity. This crossover is an universal feature of the RNiC$_2$ family and can be attributed both to the reconstruction of the conduction bands driven by magnetoelastic modification of the crystallographic structure \cite{laverock_electronic_2009, murase} upon a transition to the magnetically ordered state and to the consequent destruction of CDW state resulting in release of condensed carriers. Despite the $\simeq 40\%$ resistivity drop in GdNiC$_2$, its magnitude is notably smaller than in SmNiC$_2$ \cite{prathiba_tuning_2016}.  These observations are in agreement with the X-Ray data recently collected on single crystals by Shimomura et al. \cite{Shimomura2016}, who have shown that, in contrast to complete suppression of CDW in the FM state of SmNiC$_2$ \cite{hanasaki_magnetic_2012, Shimomura_2009}, the Peierls instability, although weakened, survives in the magnetically ordered state of GdNiC$_2$. Note that the values of $\rho_0$ found by us are an order of magnitude larger than in the single crystals studied by Shimomura et al. Also, in our polycrystalline sample, $\rho_0(T)$ does not show any influence of the lock-in transition occurring at $T_{lock-in}\approx90$ K as seen in resistivity measured along $c$ axis of the single crystal. The polycrystalline nature of our sample is also responsible for a lower, in comparison to single crystal, value of $T_P$ (although the value of $T_P$ found by us converges with data reported by Murase et al. \cite{murase}). In the metallic regime above $T_P$, the magnetoresistance (MR = $\frac{\rho(H)-\rho_0}{\rho_0}$) is negligibly small. Below $T_P$, one observes a significant decrease of resistivity in the presence of magnetic fields and the MR remains large and negative even at lowest temperatures, softening the drop of $\rho(H)$ in proximity of $T_N$. An interesting observation is the occurrence of a small minimum followed by a hump at temperatures slightly below $T_N$ in the presence of fields ranging from 5 to 7 T (see inset of figure \ref{RT}). Considering the phase diagram of GdNiC$_2$ (see fig \ref{CP}), one can attribute this effect to the transition towards an intermediate magnetic phase. This effect will be more extensively discussed in the next paragraphs.

At temperatures $T_N<T<T_P$, the MR follows an $\sim-H^2$ dependence which is depicted in figure \ref{MR}. This behavior suggests that the main source of magnetoresistance in this temperature interval is the Zeeman splitting of the electronic bands at the Fermi level towards spin-up and spin-down ones separated by $\Delta E = 2 \mu_B H$. Theoretical work of Dieterich and Fulde \cite{Dieterich1973} predicted that a sufficiently strong magnetic field reduces the pairing interaction between electrons, lowers the CDW electronic gap ($\Delta_{CDW}$) and suppresses the CDW ground state. As a consequence, the Peierls temperature follows the BCS relation with magnetic field;

\begin{equation}
\label{MReq2}
\frac{T_P(H)-T_P(0)}{T_{P}(0)}=\frac{\gamma}{4}\left( \frac{\mu_BH}{k_BT_P(0)}\right)^2
\end{equation}

where $\gamma$ is a constant of the order of unity. When $\Delta_{CDW}$ is much larger than $\mu_B H$,  the negative magnetoresistance due to an increase of free electronic carriers can be expressed by the formula \cite{tiedje_magnetoresistance_1975}:

\begin{equation}
\label{Zeeman}
MR = \frac{\rho (H)-\rho_{0}}{\rho_0}= -\frac{1}{2} \left( \frac{\mu_BH}{k_BT}\right)^2+0 \left( \frac{\mu_BH}{k_BT}\right)^4
\end{equation}

\begin{figure*}[ht]
 \includegraphics[scale=0.50]{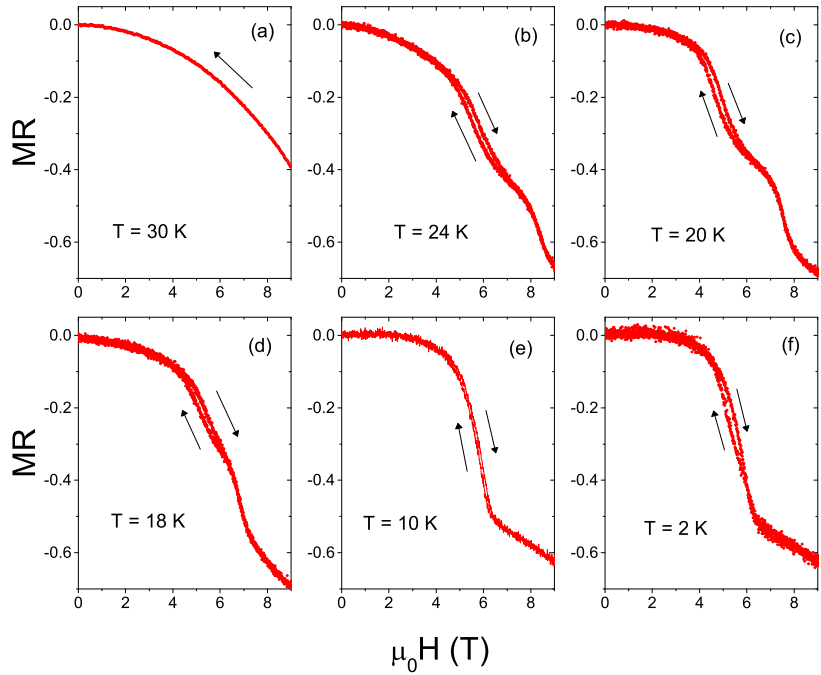}
  \caption{\label{MRLT} Magnetoresistance of GdNiC$_2$ as a function of magnetic field measured at (a) 30 K, (b) 24 K, (c) 20 K, (d) 18 K, (e) 10 K (f) 2 K. Arrows show the direction of magnetic field sweep.}
  \end{figure*}

As depicted in the inset of figure \ref{MR}, the magnetoresistance of GdNiC$_2$ scales with  $ -\left( \frac{\mu_BH}{k_BT}\right)^2$ in the wide temperature interval of $30 $ K $ \leq T \leq 180 $ K. This evidences that Zeeman suppression of the CDW is a driving force of large negative MR in this temperature range. An interesting observation, is that the description of MR with eq. \ref{Zeeman} requires the introduction of a prefactor of $\simeq$ 30. Typically, in CDW materials this coefficient or equivalently the $\gamma$ factor in eq. \ref{MReq2} is smaller than unity. Good examples are the magnetoresistance scaled by 0.25 in Li$_{0.9}$Mo$_6$O$_{17}$ \cite{xu_directional_2009} or a number of organic compounds with $\gamma<1$ \cite{brooks_magnetic_2008, graf_suppression_2004, Graf2005, monchi_international_1999, matos_modification_1996}. 
Matos \textit{et al.}\cite{matos_modification_1996} shown that the presence of weakly magnetic chains in (Per)$_2$Pt(mnt)$_2$ leading to the local increase of internal magnetic field, enhances the CDW suppression with $H$. As a result one observes the $\gamma$ parameter larger in comparison with a similar compound (Per)$_2$Au(mnt)$_2$ in which the magnetic chains are absent. This effect is in some extent similar to the case of GdNiC$_2$. However, in contrast to the aforementioned systems, where the  local magnetic moments are insignificant, in GdNiC$_2$, the Gd$^{3+}$ ions carry a large moment of ca. $8 \mu_B$ \cite{kittel_2004}. The presence of such local moments produces a strong internal magnetic field acting on neighboring Ni atoms, and that results in the large value of magnetoresistance prefactor. Such 'transferred' magnetic fields have been observed through M\"ossbauer spectroscopy (MS) measurements in hyperfine structures of various nonmagnetic atoms embedded in magnetic systems \cite{Nowik_2012}. The MS measurements on GdNiC$_2$ at 4.2 K have revealed a hyperfine magnetic field of 34 T acting at Gd nuclei \cite{matsuo_antiferromagnetism_1996,Onodera_1997}, but no reports on Ni hyperfine fields exist to the author's knowledge. We also suggest, that due to the presence of such strong magnetic moments in GdNiC$_2$ one can safely assume, that the enhancement of the CDW suppression originates mostly from the spin magnetic moments from Gd$^{3+}$ and the contribution to $\gamma$ from the orbital effects, observed in (Per)$_2$Au(mnt)$_2$ \cite{graf_suppression_2004}, is insignificant in comparison to the influence of the spin mechanism. 

Although the magnetoresistance stands in agreement with theory, we have found no visible modification of the Peierls temperature in magnetic field. The shift of $T_P$ due to Zeeman suppression of the CDW gap has been observed only in several materials with the $T_P$ as low as 8 K or 12 K\cite{bonfait_research_1995-1}. Since the Peierls temperature in GdNiC$_2$ is relatively high, $k_BT_P$ is two orders of magnitude larger than $\mu_BH$ at our maximum field of 9 T. Then, even considering the factor $\gamma$ = 30 in eq. \ref{MReq2}, the expected $T_P$ shift is only  $\simeq$ 1.3 K at 9 T. Such a small difference is difficult to observe within experimental resolution for the polycrystalline sample. Furthermore, lack of visible deviation from the mean-field (eq. \ref{Zeeman}) scaling, including signs of saturation, which was seen in Li$_{0.9}$Mo$_6$O$_{17}$ \cite{xu_directional_2009} due to complete destruction of CDW at high magnetic fields suggests, that the 
suppression of the charge density wave observed in GdNiC$_2$ is not complete for $T_N<T<T_P$, and the Peierls instability (at least partly) survives in the presence of the external magnetic field of 9 T.

Figure \ref{MRLT} shows the magnetoresistance of GdNiC$_2$ measured at temperatures in the vicinity and below $T_N$. The curve measured at $T$ = 30 K shows $-H^2$ behavior described above. As temperature is lowered to 24 K, one can observe two kinks in MR$(H)$. First, one appears at, roughly, 4.5 T and is accompanied by a small hysteresis; a second one is found at $\simeq$ 6.5 T and shows no hysteretic behavior.  Upon further decreasing the temperature, the former term becomes sharper and dominates over the latter one. Eventually, at 10 K, MR$(H)$ shows an abrupt drop at fields between 4 and 6 T. At higher fields, MR shows a further, yet much slower, decrease. Owing to the phase diagram of GdNiC$_2$ and our magnetization data, we find that both anomalies correspond to the phase transitions towards a metamagnetic phase (MM in fig \ref{CP}b). Above 10 K this process occurs via the intermediate magnetic phase, which explains the existence of two kinks in MR($H$). In contrast to that, at T $\leq$ 10 K, the increase of magnetic field transforms the system directly from AF to metamagnetic phase without any intermediate stage, which is pronounced by a single anomaly.
This steep decrease of resistivity is reminiscent with the behavior of magnetoresistance in NdNiC$_2$ \cite{Yamamoto_2013}, where the CDW state surviving the transition to AF state was definitely suppressed by a magnetic field induced spin-flop transition. A similar effect was observed slightly above the Curie temperature in SmNiC$_2$ \cite{hanasaki_magnetic_2012}, where application of magnetic field causes the transition into a ferromagnetic state, which results in a destruction of CDW. 
This suggests, that the effect observed in GdNiC$_2$ is of the same origin as in the compounds recalled above. 

The wavevector $q$ of the CDW modulation appears to play a key role in the interplay between CDW and AF. Agreement between $q$ and the magnetic propagation vector leads to a sort of resonance, which prevents the local magnetic moments from breaking the pairing interaction and, in this scenario CDW and AF orders coexist \cite{Yamamoto_2013}. Deviation from this condition leads then to breaking of the singlet electron - hole pairs forming the CDW. Although the magnetic structure of GdNiC$_2$ has not been precisely defined yet, Matsuo et al. \cite{matsuo_antiferromagnetism_1996} proposed the propagation vector of $\left[ \frac{1}{2}; \frac{1}{2}; 0\right] $. This value corresponds with the CDW modulation vector, which becomes doubly commensurate at $T_{lock-in}$ \cite{Shimomura2016}. When a magnetic field induces a change of magnetic structure this resonance becomes disturbed, which leads to suppression of the CDW and release of condensed carriers. One shall also consider the possibility of the electronic bands structure modification upon the AF - MM transition. This leads to the change of the nesting conditions and may eventually act as another mechanism suppressing the CDW.

To support this scenario, we have followed an analysis proposed by Yamamoto et al. \cite{Yamamoto_2013} and compared the ratio of resistivity in GdNiC$_2$ measured at 2 K (thus in presence of magnetic order), with and without magnetic field, respectively and at a temperature slightly above the AF transition, with   corresponding values obtained for and NdNiC$_2$. In GdNiC$_2$, $\frac{\rho_{0T,2K}}{\rho_{0T,23K}}$ = 0.51 and $\frac{\rho_{9T,2K}}{\rho_{0T,23K}}$= 0.18~ \cite{Yamamoto_2013}. These quantities parallel relevant ratios for NdNiC$_2$: in the presence of AF, partially suppressing CDW: $\frac{R_{0T,5K}}{R_{0T,20K}} \simeq$ 0.4 and $\frac{R_{9T,5K}}{R_{0T,20K}}\simeq$ 0.16 at the same temperature, albeit in the presence of a magnetic field in which the CDW is completely suppressed. To compare, in SmNiC$_2$\cite{Shimomura_2009, prathiba_tuning_2016}, where the FM order entirely destroys the CDW, the ratio $\frac{R_{5K}}{R_{20K}}\simeq$ 0.1. These results are consistent with the scenario of partial destruction of the CDW in the AF state of GdNiC$_2$ and the further suppression of the Peierls instability with increasing magnetic field, which drives the metamagnetic crossover. We emphasize that, although this comparison suggests a strong suppression of the Peierls instability in the MM state, X-Ray diffuse scattering experiment showing the absence of satellite reflections is required to deliver  unambiguous evidence of complete CDW destruction.

\begin{figure}[h]
 \includegraphics[scale=0.32]{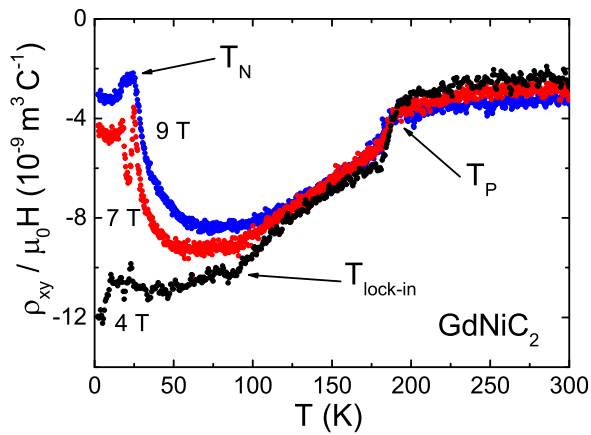}
  \caption{\label{Hall1}Hall coefficient of GdNiC$_2$ vs temperature at various fields. Arrows indicate 
characteristic phase transitions temperatures.}
\end{figure}

Due to polycrystalline nature of our samples, we were unable to explore the thermal and magnetic field evolution of CDW satellite peaks via X-Ray diffuse scattering. Instead, we have conducted a study of the Hall effect to complement the magnetoresistance data. Figure \ref{Hall1} shows the thermal dependence of the Hall resistivity normalized by magnetic field ($\frac{\rho_{xy}}{\mu_0H}$) measured at various magnetic fields. Above $T_P$, $\frac{\rho_{xy}}{\mu_0H}$ is almost temperature independent. At the Peierls temperature one observes a decrease of Hall resistivity due to condensation of part of the electronic carriers into the CDW state. The lock - in transition is pronounced as an inflection in the $\frac{\rho_{xy}}{\mu_0H}(T)$ curve. Note that,  as $T$ approaches $T_N$, one observes an increase of $\frac{\rho_{xy}}{\mu_0H}$, which is significantly enhanced at strong magnetic fields. Similar behaviors have been reported at $T_P$ and $T_C$ of SmNiC$_2$ \cite{Kim_2012}, corresponding to formation and suppression of the CDW, respectively.
Relating the increase of Hall resistivity at $T \rightarrow T_P$ exclusively to the release of free carriers due to destruction of the CDW by AF order would be too far a simplification. In materials exhibiting significant magnetic ordering, the Hall resistance consists of two components related to external magnetic field and magnetization respectively\cite{karplus_hall_1954}: 

\begin{equation}
\label{Halleq}
\rho_{xy}=R_0\mu_0H+4\pi R_SM
\end{equation}

$R_0$ is the ordinary Hall coefficient, which for a single band system, is a direct measure of electronic concentration $n$ ($R_0=\frac{1}{en}$). $R_s$ represents the anomalous Hall effect associated with skew and side jump scattering. The separation of those parameters is not straightforward and usually requires measurements in magnetic fields strong enough to observe saturation of $M(H)$ \cite{Berger1980, shiomi_extrinsic_2009, Granovskii2012}, which in an antiferromagnetic metal can be as large as tens of Teslas. Considering the $M(H)$ dependence, we can propose at least a qualitative discussion of the evolution of $R_0$ (thus of $n$) as a function of $H$.
 
 \begin{figure}[h]
 \includegraphics[scale=0.32]{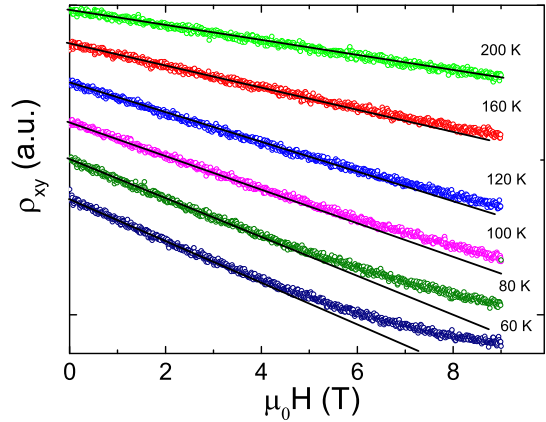}
  \caption{\label{Hall2}Hall resistance of GdNiC$_2$ as a function of H for various temperatures above 60 K. Solid lines show the extended fits to the linear parts of the curves. For clarity, the curves have been vertically shifted.}
  \end{figure}

Firstly, we have followed the Hall resistance as a function of $H$ for $T\geq60 $ K. 
The idea is that in this temperature range the magnetization is a linear function of magnetic field. Hence, the anomalous Hall resistance term is proportional to $H$ as well, and any departure from linearity of $\rho_{xy}(H)$ is a fingerprint of some modification of free carrier concentration. 
Figure \ref{Hall2} shows the Hall resistivity as a function of magnetic field for $T\geq60 $ K. At 200 K, thus above $T_P$ where GdNiC$_2$ exhibits ordinary metallic character, the Hall resistivity shows a classical linear dependence of $H$. At temperatures below $T_P$, one observes a clear deviation from this linear scaling. This effect becomes more pronounced as $T$ decreases and $H$ increases. This indicates the increase of free electronic concentration and is consistent with the negative Zeeman magnetoresistance due to partial release of CDW condensed electrons, observed in this temperature range.

 \begin{figure}[h]
  \includegraphics[scale=0.32]{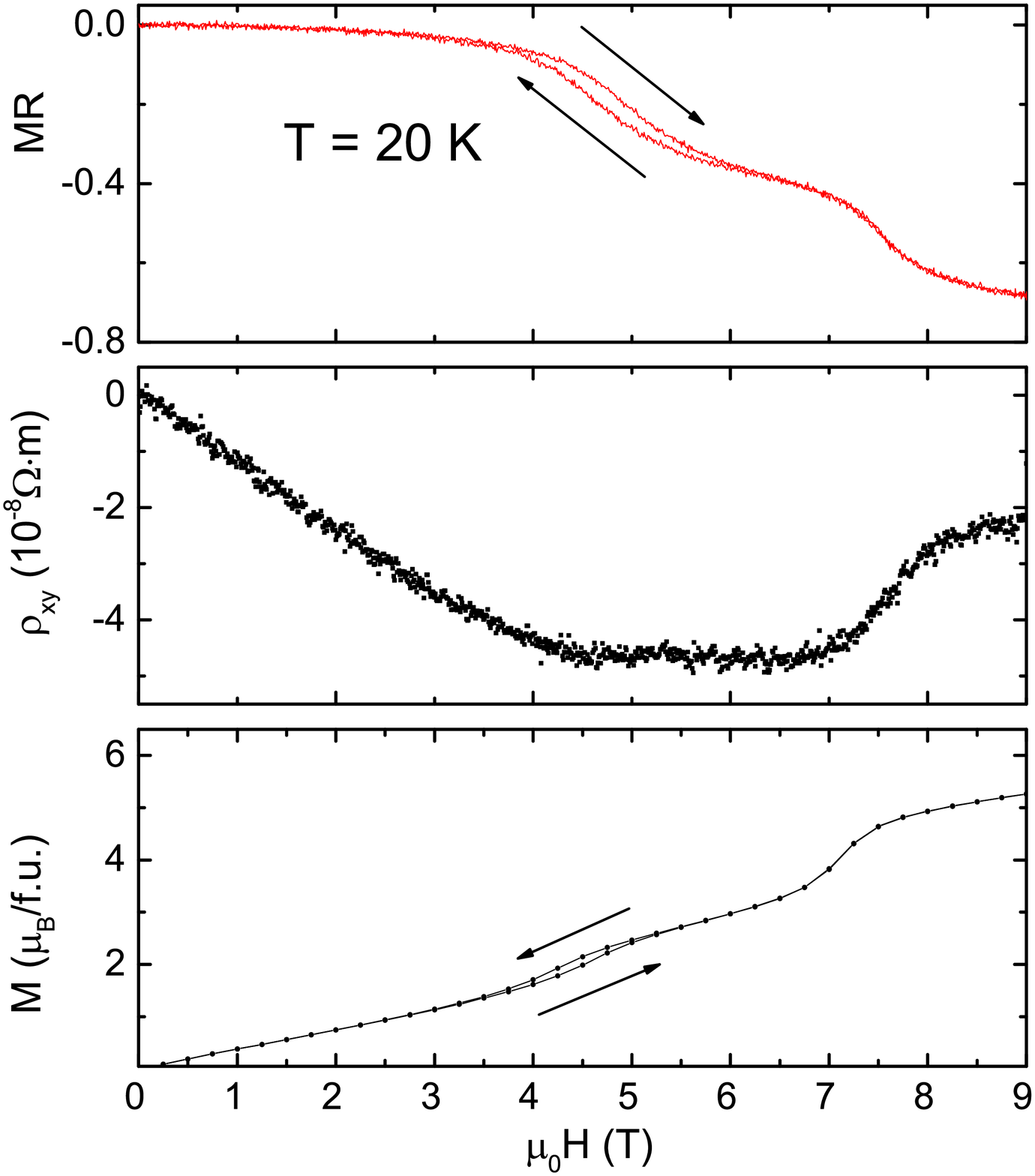}
 \caption{\label{por20}  Comparison of magnetoresistance, Hall resistivity and magnetization of GdNiC$_2$ at $T$ = 20 K. Arrows in the upper and lower panels show the direction of the magnetic field sweep. }
 \end{figure}  
 
\begin{figure}[h]
  \includegraphics[scale=0.32]{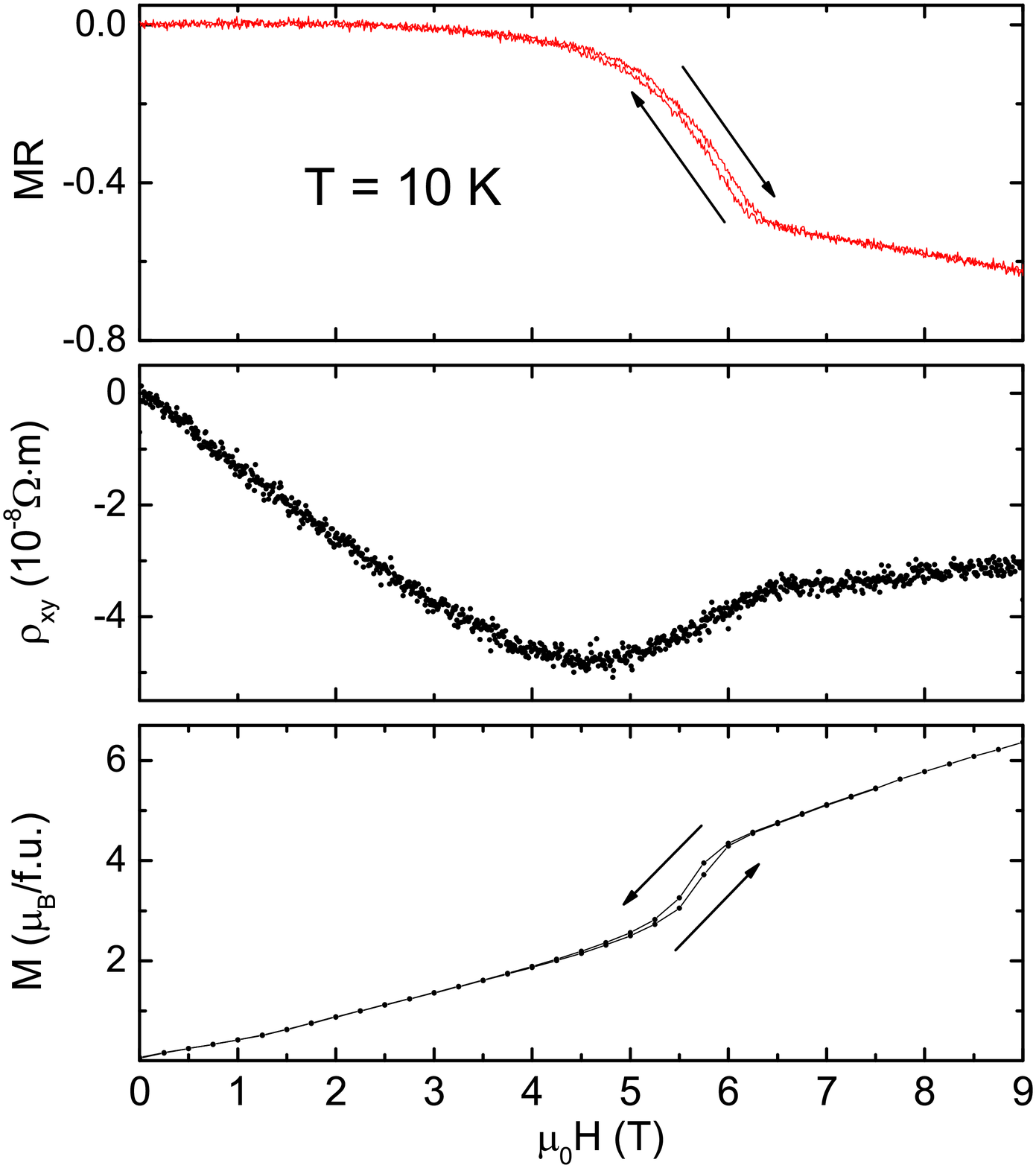}
 \caption{\label{por10} Comparison of magnetoresistance, Hall resistivity and magnetization of GdNiC$_2$ at $T$ = 10 K. Arrows in the upper and lower panels show the direction of the magnetic field sweep. }
 \end{figure}

 \begin{figure}[h]
  \includegraphics[scale=0.32]{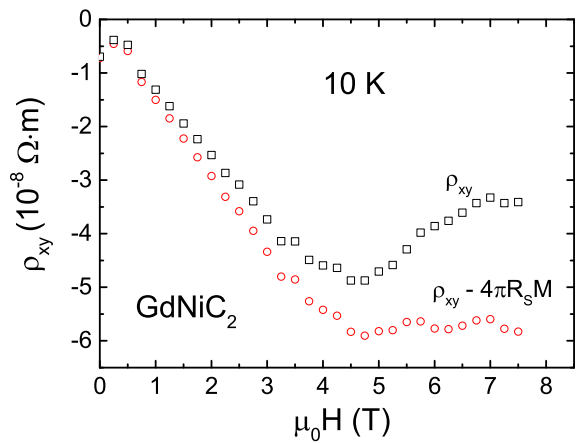}
 \caption{\label{rhest} Hall resistivity of GdNiC$_2$ measured at 10 K (black squares) and $\rho_{xy}$ after subtraction of the estimated anomalous component (red circles) - see text for details}
 \end{figure}
 
Figures \ref{por20} and \ref{por10} compare of magnetoresistance, Hall resistivity and magnetization of GdNiC$_2$ measured at 20 and 10 K, respectively. In both cases, at fields up to 3.5 T, the Hall resistance follows a linear dependence on $H$. At 4 T, there is an upturn of $\rho_{xy}(H)$ concomitant with a decrease of resistance and increase of magnetization discussed in previous paragraphs. At 20 K, the second kink observed in MR and $M$ is also reflected in $\rho_{xy}$ and is pronounced as an upturn of Hall resistivity. The departure of $\rho_{xy}$ from its linear field dependence show large similarities with the data collected at $T\geq$ 60 K (see figure \ref{Hall2}), where the carrier concentration is increased due to partial suppression of the CDW. Nevertheless, to avoid  overinterpretation of this result one has to analyze the data also in respect to the anomalous part of the Hall effect.

Due to the complicated shapes of $\rho_{xy}(H)$ and $M(H)$, the separation of ordinary and anomalous components of the Hall resistance requires several crude assumptions, which cause the approximate nature of the following analysis. Firstly, we assume that $R_S$ does not change appreciably  across the observed magnetic phase transitions. A second assumption is that at high fields $R_0\mu_0H\ll4\pi R_SM$. Then, equation \ref{Halleq} reduces to $\rho_{xy}=4\pi R_SM$. The anomalous Hall coefficient $R_S$ was evaluated from a linear fit (not shown here) to $\rho_{xy}(M)$ measured at 10 K. Then, the anomalous component was subtracted from the measured $\rho_{xy}(H)$. The result is shown in fig. \ref{rhest}. A clear deviation of this curve from a linear function of H suggests the increase of the carrier concentration at field stronger than 4 T. This supports the scenario of partial CDW destruction due to a magnetic transition.

\section{conclusions}
We have studied the specific heat, magnetic, magnetotransport and galvanomagnetic properties of polycrystalline GdNiC$_2$. In the absence of antiferromagnetic order, above $T_N$, we observe a strong negative magnetoresistance due to Zeeman splitting of conduction bands and partial suppression of the CDW. This result is confirmed by the increase of electronic carrier concentration revealed by Hall measurements. The presence of large local magnetic moments of Gd$^{3+}$ ions is presumably responsible for the anomalously strong magnetic field dependence of magnetoresistance. In order to investigate this problem more deeply, a measurement of local magnetic field, especially acting on Ni atoms, should be performed using a local-probe technique, like M\"ossbauer spectroscopy.
At temperatures below $T_N$, we observe a significant decrease of resistance as magnetic field drives the crossover from aniferromagnetic to metamagnetic order. We suggest that the evolution of the magnetic propagation vector upon the MM transition distorts the resonance between AF magnetic order and doubly commensurate CDW. As a result the electron - hole pairing interaction is substantially weakened.    This, together with possible modification of the Fermi surface nesting conditions, results in strong suppression of the charge density wave instability in GdNiC$_2$ and a prominent release of electronic carriers, which we confirm by magnetoresistance and Hall effect measurements.
We also suggest, that an X-Ray study of structural modulation response to the application of magnetic fields performed on a single crystal is necessary to investigate further the behavior of the CDW ground state in this strongly magnetic system.
\\
\section{Acknowledgments}
We would like to thank to Joe D. Thompson (LANL) for helpful discussions. Authors gratefully acknowledge the financial support from National Science Centre (Poland), grant number:  UMO-2015/19/B/ST3/03127.

\begin{thebibliography}{55}%
\makeatletter
\providecommand \@ifxundefined [1]{%
 \@ifx{#1\undefined}
}%
\providecommand \@ifnum [1]{%
 \ifnum #1\expandafter \@firstoftwo
 \else \expandafter \@secondoftwo
 \fi
}%
\providecommand \@ifx [1]{%
 \ifx #1\expandafter \@firstoftwo
 \else \expandafter \@secondoftwo
 \fi
}%
\providecommand \natexlab [1]{#1}%
\providecommand \enquote  [1]{``#1''}%
\providecommand \bibnamefont  [1]{#1}%
\providecommand \bibfnamefont [1]{#1}%
\providecommand \citenamefont [1]{#1}%
\providecommand \href@noop [0]{\@secondoftwo}%
\providecommand \href [0]{\begingroup \@sanitize@url \@href}%
\providecommand \@href[1]{\@@startlink{#1}\@@href}%
\providecommand \@@href[1]{\endgroup#1\@@endlink}%
\providecommand \@sanitize@url [0]{\catcode `\\12\catcode `\$12\catcode
  `\&12\catcode `\#12\catcode `\^12\catcode `\_12\catcode `\%12\relax}%
\providecommand \@@startlink[1]{}%
\providecommand \@@endlink[0]{}%
\providecommand \url  [0]{\begingroup\@sanitize@url \@url }%
\providecommand \@url [1]{\endgroup\@href {#1}{\urlprefix }}%
\providecommand \urlprefix  [0]{URL }%
\providecommand \Eprint [0]{\href }%
\providecommand \doibase [0]{http://dx.doi.org/}%
\providecommand \selectlanguage [0]{\@gobble}%
\providecommand \bibinfo  [0]{\@secondoftwo}%
\providecommand \bibfield  [0]{\@secondoftwo}%
\providecommand \translation [1]{[#1]}%
\providecommand \BibitemOpen [0]{}%
\providecommand \bibitemStop [0]{}%
\providecommand \bibitemNoStop [0]{.\EOS\space}%
\providecommand \EOS [0]{\spacefactor3000\relax}%
\providecommand \BibitemShut  [1]{\csname bibitem#1\endcsname}%
\let\auto@bib@innerbib\@empty
\bibitem [{\citenamefont {Grüner}(1994)}]{gruner_density_1994}%
  \BibitemOpen
  \bibfield  {author} {\bibinfo {author} {\bibfnamefont {G.}~\bibnamefont
  {Grüner}},\ }\href@noop {} {\emph {\bibinfo {title} {Density {Waves} in
  {Soilds}, {Frontiers} in {Physicvs}}}},\ Vol.~\bibinfo {volume} {89}\
  (\bibinfo  {publisher} {Addison-Wesley},\ \bibinfo {address} {New York},\
  \bibinfo {year} {1994})\BibitemShut {NoStop}%
\bibitem [{\citenamefont {Monceau}(2012)}]{Monceau2012}%
  \BibitemOpen
  \bibfield  {author} {\bibinfo {author} {\bibfnamefont {P.}~\bibnamefont
  {Monceau}},\ }\href {\doibase 10.1080/00018732.2012.719674} {\bibfield
  {journal} {\bibinfo  {journal} {Advances in Physics}\ }\textbf {\bibinfo
  {volume} {61}},\ \bibinfo {pages} {325} (\bibinfo {year} {2012})}\BibitemShut
  {NoStop}%
\bibitem [{\citenamefont {Gr\"uner}(1988)}]{Gruner1988}%
  \BibitemOpen
  \bibfield  {author} {\bibinfo {author} {\bibfnamefont {G.}~\bibnamefont
  {Gr\"uner}},\ }\href {\doibase 10.1103/RevModPhys.60.1129} {\bibfield
  {journal} {\bibinfo  {journal} {Rev. Mod. Phys.}\ }\textbf {\bibinfo {volume}
  {60}},\ \bibinfo {pages} {1129} (\bibinfo {year} {1988})}\BibitemShut
  {NoStop}%
\bibitem [{\citenamefont {Tiedje}\ \emph {et~al.}(1975)\citenamefont {Tiedje},
  \citenamefont {Carolan}, \citenamefont {Berlinsky},\ and\ \citenamefont
  {Weiler}}]{tiedje_magnetoresistance_1975}%
  \BibitemOpen
  \bibfield  {author} {\bibinfo {author} {\bibfnamefont {T.}~\bibnamefont
  {Tiedje}}, \bibinfo {author} {\bibfnamefont {J.~F.}\ \bibnamefont {Carolan}},
  \bibinfo {author} {\bibfnamefont {A.~J.}\ \bibnamefont {Berlinsky}}, \ and\
  \bibinfo {author} {\bibfnamefont {L.}~\bibnamefont {Weiler}},\ }\href
  {\doibase 10.1139/p75-202} {\bibfield  {journal} {\bibinfo  {journal}
  {Canadian Journal of Physics}\ }\textbf {\bibinfo {volume} {53}},\ \bibinfo
  {pages} {1593} (\bibinfo {year} {1975})}\BibitemShut {NoStop}%
\bibitem [{\citenamefont {Balseiro}\ and\ \citenamefont
  {Falicov}(1985)}]{Balseiro1985}%
  \BibitemOpen
  \bibfield  {author} {\bibinfo {author} {\bibfnamefont {C.~A.}\ \bibnamefont
  {Balseiro}}\ and\ \bibinfo {author} {\bibfnamefont {L.~M.}\ \bibnamefont
  {Falicov}},\ }\href {\doibase 10.1103/PhysRevLett.55.2336} {\bibfield
  {journal} {\bibinfo  {journal} {Phys. Rev. Lett.}\ }\textbf {\bibinfo
  {volume} {55}},\ \bibinfo {pages} {2336} (\bibinfo {year}
  {1985})}\BibitemShut {NoStop}%
\bibitem [{\citenamefont {Brooks}\ \emph {et~al.}(2006)\citenamefont {Brooks},
  \citenamefont {Graf}, \citenamefont {Choi}, \citenamefont {Almeida},
  \citenamefont {Dias}, \citenamefont {Henriques},\ and\ \citenamefont
  {Matos}}]{brooks_magnetic_2006}%
  \BibitemOpen
  \bibfield  {author} {\bibinfo {author} {\bibfnamefont {J.~S.}\ \bibnamefont
  {Brooks}}, \bibinfo {author} {\bibfnamefont {D.}~\bibnamefont {Graf}},
  \bibinfo {author} {\bibfnamefont {E.~S.}\ \bibnamefont {Choi}}, \bibinfo
  {author} {\bibfnamefont {M.}~\bibnamefont {Almeida}}, \bibinfo {author}
  {\bibfnamefont {J.~C.}\ \bibnamefont {Dias}}, \bibinfo {author}
  {\bibfnamefont {R.~T.}\ \bibnamefont {Henriques}}, \ and\ \bibinfo {author}
  {\bibfnamefont {M.}~\bibnamefont {Matos}},\ }\href {\doibase
  10.1016/j.cap.2005.01.039} {\bibfield  {journal} {\bibinfo  {journal}
  {Current Applied Physics}\ }\textbf {\bibinfo {volume} {6}},\ \bibinfo
  {pages} {913} (\bibinfo {year} {2006})}\BibitemShut {NoStop}%
\bibitem [{\citenamefont {Zanchi}\ \emph {et~al.}(1996)\citenamefont {Zanchi},
  \citenamefont {Bjeliš},\ and\ \citenamefont
  {Montambaux}}]{zanchi_phase_1996}%
  \BibitemOpen
  \bibfield  {author} {\bibinfo {author} {\bibfnamefont {D.}~\bibnamefont
  {Zanchi}}, \bibinfo {author} {\bibfnamefont {A.}~\bibnamefont {Bjeliš}}, \
  and\ \bibinfo {author} {\bibfnamefont {G.}~\bibnamefont {Montambaux}},\
  }\href {\doibase 10.1103/PhysRevB.53.1240} {\bibfield  {journal} {\bibinfo
  {journal} {Physical Review B}\ }\textbf {\bibinfo {volume} {53}},\ \bibinfo
  {pages} {1240} (\bibinfo {year} {1996})}\BibitemShut {NoStop}%
\bibitem [{\citenamefont {Graf}\ \emph
  {et~al.}(2004{\natexlab{a}})\citenamefont {Graf}, \citenamefont {Choi},
  \citenamefont {Brooks}, \citenamefont {Matos}, \citenamefont {Henriques},\
  and\ \citenamefont {Almeida}}]{graf2004}%
  \BibitemOpen
  \bibfield  {author} {\bibinfo {author} {\bibfnamefont {D.}~\bibnamefont
  {Graf}}, \bibinfo {author} {\bibfnamefont {E.~S.}\ \bibnamefont {Choi}},
  \bibinfo {author} {\bibfnamefont {J.~S.}\ \bibnamefont {Brooks}}, \bibinfo
  {author} {\bibfnamefont {M.}~\bibnamefont {Matos}}, \bibinfo {author}
  {\bibfnamefont {R.~T.}\ \bibnamefont {Henriques}}, \ and\ \bibinfo {author}
  {\bibfnamefont {M.}~\bibnamefont {Almeida}},\ }\href {\doibase
  10.1103/PhysRevLett.93.076406} {\bibfield  {journal} {\bibinfo  {journal}
  {Phys. Rev. Lett.}\ }\textbf {\bibinfo {volume} {93}},\ \bibinfo {pages}
  {076406} (\bibinfo {year} {2004}{\natexlab{a}})}\BibitemShut {NoStop}%
\bibitem [{\citenamefont {Young}\ and\ \citenamefont
  {Sokoloff}(1974)}]{Young1974}%
  \BibitemOpen
  \bibfield  {author} {\bibinfo {author} {\bibfnamefont {C.~Y.}\ \bibnamefont
  {Young}}\ and\ \bibinfo {author} {\bibfnamefont {J.~B.}\ \bibnamefont
  {Sokoloff}},\ }\href {http://stacks.iop.org/0305-4608/4/i=8/a=023} {\bibfield
   {journal} {\bibinfo  {journal} {Journal of Physics F: Metal Physics}\
  }\textbf {\bibinfo {volume} {4}},\ \bibinfo {pages} {1304} (\bibinfo {year}
  {1974})}\BibitemShut {NoStop}%
\bibitem [{\citenamefont {Fawcett}(1988)}]{Fawcett1988}%
  \BibitemOpen
  \bibfield  {author} {\bibinfo {author} {\bibfnamefont {E.}~\bibnamefont
  {Fawcett}},\ }\href {\doibase 10.1103/RevModPhys.60.209} {\bibfield
  {journal} {\bibinfo  {journal} {Rev. Mod. Phys.}\ }\textbf {\bibinfo {volume}
  {60}},\ \bibinfo {pages} {209} (\bibinfo {year} {1988})}\BibitemShut
  {NoStop}%
\bibitem [{\citenamefont {Jacques}\ \emph {et~al.}(2014)\citenamefont
  {Jacques}, \citenamefont {Pinsolle}, \citenamefont {Ravy}, \citenamefont
  {Abramovici},\ and\ \citenamefont {Le~Bolloc'h}}]{jacques2014}%
  \BibitemOpen
  \bibfield  {author} {\bibinfo {author} {\bibfnamefont {V.~L.~R.}\
  \bibnamefont {Jacques}}, \bibinfo {author} {\bibfnamefont {E.}~\bibnamefont
  {Pinsolle}}, \bibinfo {author} {\bibfnamefont {S.}~\bibnamefont {Ravy}},
  \bibinfo {author} {\bibfnamefont {G.}~\bibnamefont {Abramovici}}, \ and\
  \bibinfo {author} {\bibfnamefont {D.}~\bibnamefont {Le~Bolloc'h}},\ }\href
  {\doibase 10.1103/PhysRevB.89.245127} {\bibfield  {journal} {\bibinfo
  {journal} {Phys. Rev. B}\ }\textbf {\bibinfo {volume} {89}},\ \bibinfo
  {pages} {245127} (\bibinfo {year} {2014})}\BibitemShut {NoStop}%
\bibitem [{\citenamefont {Nicholson}\ \emph {et~al.}(2016)\citenamefont
  {Nicholson}, \citenamefont {Monney}, \citenamefont {Carley}, \citenamefont
  {Frietsch}, \citenamefont {Bowlan}, \citenamefont {Weinelt},\ and\
  \citenamefont {Wolf}}]{nicholson2016}%
  \BibitemOpen
  \bibfield  {author} {\bibinfo {author} {\bibfnamefont {C.~W.}\ \bibnamefont
  {Nicholson}}, \bibinfo {author} {\bibfnamefont {C.}~\bibnamefont {Monney}},
  \bibinfo {author} {\bibfnamefont {R.}~\bibnamefont {Carley}}, \bibinfo
  {author} {\bibfnamefont {B.}~\bibnamefont {Frietsch}}, \bibinfo {author}
  {\bibfnamefont {J.}~\bibnamefont {Bowlan}}, \bibinfo {author} {\bibfnamefont
  {M.}~\bibnamefont {Weinelt}}, \ and\ \bibinfo {author} {\bibfnamefont
  {M.}~\bibnamefont {Wolf}},\ }\href {\doibase 10.1103/PhysRevLett.117.136801}
  {\bibfield  {journal} {\bibinfo  {journal} {Phys. Rev. Lett.}\ }\textbf
  {\bibinfo {volume} {117}},\ \bibinfo {pages} {136801} (\bibinfo {year}
  {2016})}\BibitemShut {NoStop}%
\bibitem [{\citenamefont {Jacques}\ \emph {et~al.}(2016)\citenamefont
  {Jacques}, \citenamefont {Laulh\'e}, \citenamefont {Moisan}, \citenamefont
  {Ravy},\ and\ \citenamefont {Le~Bolloc'h}}]{jacques2016}%
  \BibitemOpen
  \bibfield  {author} {\bibinfo {author} {\bibfnamefont {V.~L.~R.}\
  \bibnamefont {Jacques}}, \bibinfo {author} {\bibfnamefont {C.}~\bibnamefont
  {Laulh\'e}}, \bibinfo {author} {\bibfnamefont {N.}~\bibnamefont {Moisan}},
  \bibinfo {author} {\bibfnamefont {S.}~\bibnamefont {Ravy}}, \ and\ \bibinfo
  {author} {\bibfnamefont {D.}~\bibnamefont {Le~Bolloc'h}},\ }\href {\doibase
  10.1103/PhysRevLett.117.156401} {\bibfield  {journal} {\bibinfo  {journal}
  {Phys. Rev. Lett.}\ }\textbf {\bibinfo {volume} {117}},\ \bibinfo {pages}
  {156401} (\bibinfo {year} {2016})}\BibitemShut {NoStop}%
\bibitem [{\citenamefont {Yang}\ \emph {et~al.}(1991)\citenamefont {Yang},
  \citenamefont {Klavins},\ and\ \citenamefont {Shelton}}]{Yang1991}%
  \BibitemOpen
  \bibfield  {author} {\bibinfo {author} {\bibfnamefont {H.~D.}\ \bibnamefont
  {Yang}}, \bibinfo {author} {\bibfnamefont {P.}~\bibnamefont {Klavins}}, \
  and\ \bibinfo {author} {\bibfnamefont {R.~N.}\ \bibnamefont {Shelton}},\
  }\href {\doibase 10.1103/PhysRevB.43.7688} {\bibfield  {journal} {\bibinfo
  {journal} {Phys. Rev. B}\ }\textbf {\bibinfo {volume} {43}},\ \bibinfo
  {pages} {7688} (\bibinfo {year} {1991})}\BibitemShut {NoStop}%
\bibitem [{\citenamefont {Ghosh}\ \emph {et~al.}(1993)\citenamefont {Ghosh},
  \citenamefont {Ramakrishnan},\ and\ \citenamefont {Chandra}}]{Ghosh1993}%
  \BibitemOpen
  \bibfield  {author} {\bibinfo {author} {\bibfnamefont {K.}~\bibnamefont
  {Ghosh}}, \bibinfo {author} {\bibfnamefont {S.}~\bibnamefont {Ramakrishnan}},
  \ and\ \bibinfo {author} {\bibfnamefont {G.}~\bibnamefont {Chandra}},\ }\href
  {\doibase 10.1103/PhysRevB.48.4152} {\bibfield  {journal} {\bibinfo
  {journal} {Phys. Rev. B}\ }\textbf {\bibinfo {volume} {48}},\ \bibinfo
  {pages} {4152} (\bibinfo {year} {1993})}\BibitemShut {NoStop}%
\bibitem [{\citenamefont {Galli}\ \emph {et~al.}(2000)\citenamefont {Galli},
  \citenamefont {Ramakrishnan}, \citenamefont {Taniguchi}, \citenamefont
  {Nieuwenhuys}, \citenamefont {Mydosh}, \citenamefont {Geupel}, \citenamefont
  {Lüdecke},\ and\ \citenamefont {van
  Smaalen}}]{galli_charge-density-wave_2000}%
  \BibitemOpen
  \bibfield  {author} {\bibinfo {author} {\bibfnamefont {F.}~\bibnamefont
  {Galli}}, \bibinfo {author} {\bibfnamefont {S.}~\bibnamefont {Ramakrishnan}},
  \bibinfo {author} {\bibfnamefont {T.}~\bibnamefont {Taniguchi}}, \bibinfo
  {author} {\bibfnamefont {G.~J.}\ \bibnamefont {Nieuwenhuys}}, \bibinfo
  {author} {\bibfnamefont {J.~A.}\ \bibnamefont {Mydosh}}, \bibinfo {author}
  {\bibfnamefont {S.}~\bibnamefont {Geupel}}, \bibinfo {author} {\bibfnamefont
  {J.}~\bibnamefont {Lüdecke}}, \ and\ \bibinfo {author} {\bibfnamefont
  {S.}~\bibnamefont {van Smaalen}},\ }\href {\doibase
  10.1103/PhysRevLett.85.158} {\bibfield  {journal} {\bibinfo  {journal}
  {Physical Review Letters}\ }\textbf {\bibinfo {volume} {85}},\ \bibinfo
  {pages} {158} (\bibinfo {year} {2000})}\BibitemShut {NoStop}%
\bibitem [{\citenamefont {Galli}\ \emph {et~al.}(2002)\citenamefont {Galli},
  \citenamefont {Feyerherm}, \citenamefont {Hendrikx}, \citenamefont {Dudzik},
  \citenamefont {Nieuwenhuys}, \citenamefont {Ramakrishnan}, \citenamefont
  {Brown}, \citenamefont {Smaalen},\ and\ \citenamefont
  {Mydosh}}]{galli_coexistence_2002}%
  \BibitemOpen
  \bibfield  {author} {\bibinfo {author} {\bibfnamefont {F.}~\bibnamefont
  {Galli}}, \bibinfo {author} {\bibfnamefont {R.}~\bibnamefont {Feyerherm}},
  \bibinfo {author} {\bibfnamefont {R.~W.~A.}\ \bibnamefont {Hendrikx}},
  \bibinfo {author} {\bibfnamefont {E.}~\bibnamefont {Dudzik}}, \bibinfo
  {author} {\bibfnamefont {G.~J.}\ \bibnamefont {Nieuwenhuys}}, \bibinfo
  {author} {\bibfnamefont {S.}~\bibnamefont {Ramakrishnan}}, \bibinfo {author}
  {\bibfnamefont {S.~D.}\ \bibnamefont {Brown}}, \bibinfo {author}
  {\bibfnamefont {S.~v.}\ \bibnamefont {Smaalen}}, \ and\ \bibinfo {author}
  {\bibfnamefont {J.~A.}\ \bibnamefont {Mydosh}},\ }\href {\doibase
  10.1088/0953-8984/14/20/302} {\bibfield  {journal} {\bibinfo  {journal}
  {Journal of Physics: Condensed Matter}\ }\textbf {\bibinfo {volume} {14}},\
  \bibinfo {pages} {5067} (\bibinfo {year} {2002})}\BibitemShut {NoStop}%
\bibitem [{\citenamefont {Hossain}\ \emph {et~al.}(2005)\citenamefont
  {Hossain}, \citenamefont {Schmidt}, \citenamefont {Schnelle}, \citenamefont
  {Jeevan}, \citenamefont {Geibel}, \citenamefont {Ramakrishnan}, \citenamefont
  {Mydosh},\ and\ \citenamefont {Grin}}]{Hossain2005}%
  \BibitemOpen
  \bibfield  {author} {\bibinfo {author} {\bibfnamefont {Z.}~\bibnamefont
  {Hossain}}, \bibinfo {author} {\bibfnamefont {M.}~\bibnamefont {Schmidt}},
  \bibinfo {author} {\bibfnamefont {W.}~\bibnamefont {Schnelle}}, \bibinfo
  {author} {\bibfnamefont {H.~S.}\ \bibnamefont {Jeevan}}, \bibinfo {author}
  {\bibfnamefont {C.}~\bibnamefont {Geibel}}, \bibinfo {author} {\bibfnamefont
  {S.}~\bibnamefont {Ramakrishnan}}, \bibinfo {author} {\bibfnamefont {J.~A.}\
  \bibnamefont {Mydosh}}, \ and\ \bibinfo {author} {\bibfnamefont
  {Y.}~\bibnamefont {Grin}},\ }\href {\doibase 10.1103/PhysRevB.71.060406}
  {\bibfield  {journal} {\bibinfo  {journal} {Phys. Rev. B}\ }\textbf {\bibinfo
  {volume} {71}},\ \bibinfo {pages} {060406} (\bibinfo {year}
  {2005})}\BibitemShut {NoStop}%
\bibitem [{\citenamefont {van Smaalen}\ \emph {et~al.}(2004)\citenamefont {van
  Smaalen}, \citenamefont {Shaz}, \citenamefont {Palatinus}, \citenamefont
  {Daniels}, \citenamefont {Galli}, \citenamefont {Nieuwenhuys},\ and\
  \citenamefont {Mydosh}}]{vansmaalen2004}%
  \BibitemOpen
  \bibfield  {author} {\bibinfo {author} {\bibfnamefont {S.}~\bibnamefont {van
  Smaalen}}, \bibinfo {author} {\bibfnamefont {M.}~\bibnamefont {Shaz}},
  \bibinfo {author} {\bibfnamefont {L.}~\bibnamefont {Palatinus}}, \bibinfo
  {author} {\bibfnamefont {P.}~\bibnamefont {Daniels}}, \bibinfo {author}
  {\bibfnamefont {F.}~\bibnamefont {Galli}}, \bibinfo {author} {\bibfnamefont
  {G.~J.}\ \bibnamefont {Nieuwenhuys}}, \ and\ \bibinfo {author} {\bibfnamefont
  {J.~A.}\ \bibnamefont {Mydosh}},\ }\href {\doibase
  10.1103/PhysRevB.69.014103} {\bibfield  {journal} {\bibinfo  {journal} {Phys.
  Rev. B}\ }\textbf {\bibinfo {volume} {69}},\ \bibinfo {pages} {014103}
  (\bibinfo {year} {2004})}\BibitemShut {NoStop}%
\bibitem [{\citenamefont {Hanasaki}\ \emph {et~al.}(2012)\citenamefont
  {Hanasaki}, \citenamefont {Nogami}, \citenamefont {Kakinuma}, \citenamefont
  {Shimomura}, \citenamefont {Kosaka},\ and\ \citenamefont
  {Onodera}}]{hanasaki_magnetic_2012}%
  \BibitemOpen
  \bibfield  {author} {\bibinfo {author} {\bibfnamefont {N.}~\bibnamefont
  {Hanasaki}}, \bibinfo {author} {\bibfnamefont {Y.}~\bibnamefont {Nogami}},
  \bibinfo {author} {\bibfnamefont {M.}~\bibnamefont {Kakinuma}}, \bibinfo
  {author} {\bibfnamefont {S.}~\bibnamefont {Shimomura}}, \bibinfo {author}
  {\bibfnamefont {M.}~\bibnamefont {Kosaka}}, \ and\ \bibinfo {author}
  {\bibfnamefont {H.}~\bibnamefont {Onodera}},\ }\href {\doibase
  10.1103/PhysRevB.85.092402} {\bibfield  {journal} {\bibinfo  {journal}
  {Physical Review B}\ }\textbf {\bibinfo {volume} {85}},\ \bibinfo {pages}
  {092402} (\bibinfo {year} {2012})}\BibitemShut {NoStop}%
\bibitem [{\citenamefont {Yamamoto}\ \emph {et~al.}(2013)\citenamefont
  {Yamamoto}, \citenamefont {Kondo}, \citenamefont {Maeda},\ and\ \citenamefont
  {Nogami}}]{Yamamoto_2013}%
  \BibitemOpen
  \bibfield  {author} {\bibinfo {author} {\bibfnamefont {N.}~\bibnamefont
  {Yamamoto}}, \bibinfo {author} {\bibfnamefont {R.}~\bibnamefont {Kondo}},
  \bibinfo {author} {\bibfnamefont {H.}~\bibnamefont {Maeda}}, \ and\ \bibinfo
  {author} {\bibfnamefont {Y.}~\bibnamefont {Nogami}},\ }\href {\doibase
  10.7566/JPSJ.82.123701} {\bibfield  {journal} {\bibinfo  {journal} {Journal
  of the Physical Society of Japan}\ }\textbf {\bibinfo {volume} {82}},\
  \bibinfo {pages} {123701} (\bibinfo {year} {2013})}\BibitemShut {NoStop}%
\bibitem [{\citenamefont {Shimomura}\ \emph {et~al.}(2016)\citenamefont
  {Shimomura}, \citenamefont {Hayashi}, \citenamefont {Hanasaki}, \citenamefont
  {Ohnuma}, \citenamefont {Kobayashi}, \citenamefont {Nakao}, \citenamefont
  {Mizumaki},\ and\ \citenamefont {Onodera}}]{Shimomura2016}%
  \BibitemOpen
  \bibfield  {author} {\bibinfo {author} {\bibfnamefont {S.}~\bibnamefont
  {Shimomura}}, \bibinfo {author} {\bibfnamefont {C.}~\bibnamefont {Hayashi}},
  \bibinfo {author} {\bibfnamefont {N.}~\bibnamefont {Hanasaki}}, \bibinfo
  {author} {\bibfnamefont {K.}~\bibnamefont {Ohnuma}}, \bibinfo {author}
  {\bibfnamefont {Y.}~\bibnamefont {Kobayashi}}, \bibinfo {author}
  {\bibfnamefont {H.}~\bibnamefont {Nakao}}, \bibinfo {author} {\bibfnamefont
  {M.}~\bibnamefont {Mizumaki}}, \ and\ \bibinfo {author} {\bibfnamefont
  {H.}~\bibnamefont {Onodera}},\ }\href {\doibase 10.1103/PhysRevB.93.165108}
  {\bibfield  {journal} {\bibinfo  {journal} {Physical Review B}\ }\textbf
  {\bibinfo {volume} {93}},\ \bibinfo {pages} {165108} (\bibinfo {year}
  {2016})}\BibitemShut {NoStop}%
\bibitem [{\citenamefont {{Bodak, O. I.}}\ and\ \citenamefont {{Marusin, E.
  P.}}(1979)}]{bodak_o._i.__1979}%
  \BibitemOpen
  \bibfield  {author} {\bibinfo {author} {\bibnamefont {{Bodak, O. I.}}}\ and\
  \bibinfo {author} {\bibnamefont {{Marusin, E. P.}}},\ }\href@noop {}
  {\bibfield  {journal} {\bibinfo  {journal} {Dopov. Akad. Nauk Ukr. SSR}\
  }\textbf {\bibinfo {volume} {Ser. A}},\ \bibinfo {pages} {1048} (\bibinfo
  {year} {1979})}\BibitemShut {NoStop}%
\bibitem [{\citenamefont {Jeitschko}\ and\ \citenamefont
  {Gerss}(1986)}]{jeitschko_ternary_1986}%
  \BibitemOpen
  \bibfield  {author} {\bibinfo {author} {\bibfnamefont {W.}~\bibnamefont
  {Jeitschko}}\ and\ \bibinfo {author} {\bibfnamefont {M.~H.}\ \bibnamefont
  {Gerss}},\ }\href {\doibase 10.1016/0022-5088(86)90225-0} {\bibfield
  {journal} {\bibinfo  {journal} {Journal of the Less Common Metals}\ }\textbf
  {\bibinfo {volume} {116}},\ \bibinfo {pages} {147} (\bibinfo {year}
  {1986})}\BibitemShut {NoStop}%
\bibitem [{\citenamefont {Matsuo}\ \emph {et~al.}(1996)\citenamefont {Matsuo},
  \citenamefont {Onodera}, \citenamefont {Kosaka}, \citenamefont {Kobayashi},
  \citenamefont {Ohashi}, \citenamefont {Yamauchi},\ and\ \citenamefont
  {Yamaguchi}}]{matsuo_antiferromagnetism_1996}%
  \BibitemOpen
  \bibfield  {author} {\bibinfo {author} {\bibfnamefont {S.}~\bibnamefont
  {Matsuo}}, \bibinfo {author} {\bibfnamefont {H.}~\bibnamefont {Onodera}},
  \bibinfo {author} {\bibfnamefont {M.}~\bibnamefont {Kosaka}}, \bibinfo
  {author} {\bibfnamefont {H.}~\bibnamefont {Kobayashi}}, \bibinfo {author}
  {\bibfnamefont {M.}~\bibnamefont {Ohashi}}, \bibinfo {author} {\bibfnamefont
  {H.}~\bibnamefont {Yamauchi}}, \ and\ \bibinfo {author} {\bibfnamefont
  {Y.}~\bibnamefont {Yamaguchi}},\ }\href {\doibase
  10.1016/S0304-8853(96)01282-6} {\bibfield  {journal} {\bibinfo  {journal}
  {Journal of Magnetism and Magnetic Materials}\ }\textbf {\bibinfo {volume}
  {161}},\ \bibinfo {pages} {255} (\bibinfo {year} {1996})}\BibitemShut
  {NoStop}%
\bibitem [{\citenamefont {Onodera}\ \emph {et~al.}(1998)\citenamefont
  {Onodera}, \citenamefont {Koshikawa}, \citenamefont {Kosaka}, \citenamefont
  {Ohashi}, \citenamefont {Yamauchi},\ and\ \citenamefont
  {Yamaguchi}}]{onodera_magnetic_1998}%
  \BibitemOpen
  \bibfield  {author} {\bibinfo {author} {\bibfnamefont {H.}~\bibnamefont
  {Onodera}}, \bibinfo {author} {\bibfnamefont {Y.}~\bibnamefont {Koshikawa}},
  \bibinfo {author} {\bibfnamefont {M.}~\bibnamefont {Kosaka}}, \bibinfo
  {author} {\bibfnamefont {M.}~\bibnamefont {Ohashi}}, \bibinfo {author}
  {\bibfnamefont {H.}~\bibnamefont {Yamauchi}}, \ and\ \bibinfo {author}
  {\bibfnamefont {Y.}~\bibnamefont {Yamaguchi}},\ }\href {\doibase
  10.1016/S0304-8853(97)01011-1} {\bibfield  {journal} {\bibinfo  {journal}
  {Journal of Magnetism and Magnetic Materials}\ }\textbf {\bibinfo {volume}
  {182}},\ \bibinfo {pages} {161} (\bibinfo {year} {1998})}\BibitemShut
  {NoStop}%
\bibitem [{\citenamefont {Kotsanidis}\ \emph {et~al.}(1989)\citenamefont
  {Kotsanidis}, \citenamefont {Yakinthos},\ and\ \citenamefont
  {Gamari-Seale}}]{KOTSANIDIS_1989}%
  \BibitemOpen
  \bibfield  {author} {\bibinfo {author} {\bibfnamefont {P.}~\bibnamefont
  {Kotsanidis}}, \bibinfo {author} {\bibfnamefont {J.}~\bibnamefont
  {Yakinthos}}, \ and\ \bibinfo {author} {\bibfnamefont {E.}~\bibnamefont
  {Gamari-Seale}},\ }\href {\doibase
  http://dx.doi.org/10.1016/0022-5088(89)90096-9} {\bibfield  {journal}
  {\bibinfo  {journal} {Journal of the Less Common Metals}\ }\textbf {\bibinfo
  {volume} {152}},\ \bibinfo {pages} {287 } (\bibinfo {year}
  {1989})}\BibitemShut {NoStop}%
\bibitem [{\citenamefont {Onodera}\ \emph {et~al.}(1995)\citenamefont
  {Onodera}, \citenamefont {Ohashi}, \citenamefont {Amanai}, \citenamefont
  {Matsuo}, \citenamefont {Yamauchi}, \citenamefont {Yamaguchi}, \citenamefont
  {Funahashi},\ and\ \citenamefont {Morii}}]{ONODERA_1995}%
  \BibitemOpen
  \bibfield  {author} {\bibinfo {author} {\bibfnamefont {H.}~\bibnamefont
  {Onodera}}, \bibinfo {author} {\bibfnamefont {M.}~\bibnamefont {Ohashi}},
  \bibinfo {author} {\bibfnamefont {H.}~\bibnamefont {Amanai}}, \bibinfo
  {author} {\bibfnamefont {S.}~\bibnamefont {Matsuo}}, \bibinfo {author}
  {\bibfnamefont {H.}~\bibnamefont {Yamauchi}}, \bibinfo {author}
  {\bibfnamefont {Y.}~\bibnamefont {Yamaguchi}}, \bibinfo {author}
  {\bibfnamefont {S.}~\bibnamefont {Funahashi}}, \ and\ \bibinfo {author}
  {\bibfnamefont {Y.}~\bibnamefont {Morii}},\ }\href {\doibase
  http://dx.doi.org/10.1016/0304-8853(95)00077-1} {\bibfield  {journal}
  {\bibinfo  {journal} {Journal of Magnetism and Magnetic Materials}\ }\textbf
  {\bibinfo {volume} {149}},\ \bibinfo {pages} {287 } (\bibinfo {year}
  {1995})}\BibitemShut {NoStop}%
\bibitem [{\citenamefont {Schäfer}\ \emph {et~al.}(1997)\citenamefont
  {Schäfer}, \citenamefont {Kockelmann}, \citenamefont {Will}, \citenamefont
  {Yakinthos},\ and\ \citenamefont {Kotsanidis}}]{Schafer_1997}%
  \BibitemOpen
  \bibfield  {author} {\bibinfo {author} {\bibfnamefont {W.}~\bibnamefont
  {Schäfer}}, \bibinfo {author} {\bibfnamefont {W.}~\bibnamefont
  {Kockelmann}}, \bibinfo {author} {\bibfnamefont {G.}~\bibnamefont {Will}},
  \bibinfo {author} {\bibfnamefont {J.}~\bibnamefont {Yakinthos}}, \ and\
  \bibinfo {author} {\bibfnamefont {P.}~\bibnamefont {Kotsanidis}},\ }\href
  {\doibase http://dx.doi.org/10.1016/S0925-8388(96)02564-9} {\bibfield
  {journal} {\bibinfo  {journal} {Journal of Alloys and Compounds}\ }\textbf
  {\bibinfo {volume} {250}},\ \bibinfo {pages} {565 } (\bibinfo {year}
  {1997})}\BibitemShut {NoStop}%
\bibitem [{\citenamefont {Murase}\ \emph {et~al.}(2004)\citenamefont {Murase},
  \citenamefont {Tobo}, \citenamefont {Onodera}, \citenamefont {Hirano},
  \citenamefont {Hosaka}, \citenamefont {Shimomura},\ and\ \citenamefont
  {Wakabayashi}}]{murase}%
  \BibitemOpen
  \bibfield  {author} {\bibinfo {author} {\bibfnamefont {M.}~\bibnamefont
  {Murase}}, \bibinfo {author} {\bibfnamefont {A.}~\bibnamefont {Tobo}},
  \bibinfo {author} {\bibfnamefont {H.}~\bibnamefont {Onodera}}, \bibinfo
  {author} {\bibfnamefont {Y.}~\bibnamefont {Hirano}}, \bibinfo {author}
  {\bibfnamefont {T.}~\bibnamefont {Hosaka}}, \bibinfo {author} {\bibfnamefont
  {S.}~\bibnamefont {Shimomura}}, \ and\ \bibinfo {author} {\bibfnamefont
  {N.}~\bibnamefont {Wakabayashi}},\ }\href {\doibase 10.1143/JPSJ.73.2790}
  {\bibfield  {journal} {\bibinfo  {journal} {Journal of the Physical Society
  of Japan}\ }\textbf {\bibinfo {volume} {73}},\ \bibinfo {pages} {2790}
  (\bibinfo {year} {2004})}\BibitemShut {NoStop}%
\bibitem [{\citenamefont {Laverock}\ \emph {et~al.}(2009)\citenamefont
  {Laverock}, \citenamefont {Haynes}, \citenamefont {Utfeld},\ and\
  \citenamefont {Dugdale}}]{laverock_electronic_2009}%
  \BibitemOpen
  \bibfield  {author} {\bibinfo {author} {\bibfnamefont {J.}~\bibnamefont
  {Laverock}}, \bibinfo {author} {\bibfnamefont {T.~D.}\ \bibnamefont
  {Haynes}}, \bibinfo {author} {\bibfnamefont {C.}~\bibnamefont {Utfeld}}, \
  and\ \bibinfo {author} {\bibfnamefont {S.~B.}\ \bibnamefont {Dugdale}},\
  }\href {\doibase 10.1103/PhysRevB.80.125111} {\bibfield  {journal} {\bibinfo
  {journal} {Physical Review B}\ }\textbf {\bibinfo {volume} {80}},\ \bibinfo
  {pages} {125111} (\bibinfo {year} {2009})}\BibitemShut {NoStop}%
\bibitem [{\citenamefont {Kim}\ \emph {et~al.}(2013)\citenamefont {Kim},
  \citenamefont {Lee},\ and\ \citenamefont {Shim}}]{kim_chemical_2013}%
  \BibitemOpen
  \bibfield  {author} {\bibinfo {author} {\bibfnamefont {J.~N.}\ \bibnamefont
  {Kim}}, \bibinfo {author} {\bibfnamefont {C.}~\bibnamefont {Lee}}, \ and\
  \bibinfo {author} {\bibfnamefont {J.-H.}\ \bibnamefont {Shim}},\ }\href
  {\doibase 10.1088/1367-2630/15/12/123018} {\bibfield  {journal} {\bibinfo
  {journal} {New Journal of Physics}\ }\textbf {\bibinfo {volume} {15}},\
  \bibinfo {pages} {123018} (\bibinfo {year} {2013})}\BibitemShut {NoStop}%
\bibitem [{\citenamefont {Wölfel}\ \emph {et~al.}(2010)\citenamefont
  {Wölfel}, \citenamefont {Li}, \citenamefont {Shimomura}, \citenamefont
  {Onodera},\ and\ \citenamefont {van Smaalen}}]{wolfel_commensurate_2010}%
  \BibitemOpen
  \bibfield  {author} {\bibinfo {author} {\bibfnamefont {A.}~\bibnamefont
  {Wölfel}}, \bibinfo {author} {\bibfnamefont {L.}~\bibnamefont {Li}},
  \bibinfo {author} {\bibfnamefont {S.}~\bibnamefont {Shimomura}}, \bibinfo
  {author} {\bibfnamefont {H.}~\bibnamefont {Onodera}}, \ and\ \bibinfo
  {author} {\bibfnamefont {S.}~\bibnamefont {van Smaalen}},\ }\href {\doibase
  10.1103/PhysRevB.82.054120} {\bibfield  {journal} {\bibinfo  {journal}
  {Physical Review B}\ }\textbf {\bibinfo {volume} {82}},\ \bibinfo {pages}
  {054120} (\bibinfo {year} {2010})}\BibitemShut {NoStop}%
\bibitem [{\citenamefont {Ahmad}\ \emph {et~al.}(2015)\citenamefont {Ahmad},
  \citenamefont {Min}, \citenamefont {Min}, \citenamefont {Kimura},
  \citenamefont {Seo},\ and\ \citenamefont {Kwon}}]{ahmad_evidence_2015}%
  \BibitemOpen
  \bibfield  {author} {\bibinfo {author} {\bibfnamefont {D.}~\bibnamefont
  {Ahmad}}, \bibinfo {author} {\bibfnamefont {B.~H.}\ \bibnamefont {Min}},
  \bibinfo {author} {\bibfnamefont {G.~I.}\ \bibnamefont {Min}}, \bibinfo
  {author} {\bibfnamefont {S.-I.}\ \bibnamefont {Kimura}}, \bibinfo {author}
  {\bibfnamefont {J.}~\bibnamefont {Seo}}, \ and\ \bibinfo {author}
  {\bibfnamefont {Y.~S.}\ \bibnamefont {Kwon}},\ }\href {\doibase
  10.1002/pssb.201552005} {\bibfield  {journal} {\bibinfo  {journal} {physica
  status solidi (b)}\ }\textbf {\bibinfo {volume} {252}},\ \bibinfo {pages}
  {2662} (\bibinfo {year} {2015})}\BibitemShut {NoStop}%
\bibitem [{\citenamefont {Hanasaki}\ \emph {et~al.}(2011)\citenamefont
  {Hanasaki}, \citenamefont {Mikami}, \citenamefont {Torigoe}, \citenamefont
  {Nogami}, \citenamefont {Shimomura}, \citenamefont {Kosaka},\ and\
  \citenamefont {Onodera}}]{Hanasaki2011}%
  \BibitemOpen
  \bibfield  {author} {\bibinfo {author} {\bibfnamefont {N.}~\bibnamefont
  {Hanasaki}}, \bibinfo {author} {\bibfnamefont {K.}~\bibnamefont {Mikami}},
  \bibinfo {author} {\bibfnamefont {S.}~\bibnamefont {Torigoe}}, \bibinfo
  {author} {\bibfnamefont {Y.}~\bibnamefont {Nogami}}, \bibinfo {author}
  {\bibfnamefont {S.}~\bibnamefont {Shimomura}}, \bibinfo {author}
  {\bibfnamefont {M.}~\bibnamefont {Kosaka}}, \ and\ \bibinfo {author}
  {\bibfnamefont {H.}~\bibnamefont {Onodera}},\ }\href {\doibase
  10.1088/1742-6596/320/1/012072} {\bibfield  {journal} {\bibinfo  {journal}
  {Journal of Physics: Conference Series}\ }\textbf {\bibinfo {volume} {320}},\
  \bibinfo {pages} {012072} (\bibinfo {year} {2011})}\BibitemShut {NoStop}%
\bibitem [{PPM(2010)}]{PPMS_HC_manual}%
  \BibitemOpen
  \href@noop {} {\emph {\bibinfo {title} {Physical Property Measurement System
  Heat Capacity Option User's Manual, 1085-150, Rev. L3}}},\ \bibinfo
  {organization} {Quantum Design} (\bibinfo {year} {2010})\BibitemShut
  {NoStop}%
\bibitem [{\citenamefont {Onodera}\ \emph {et~al.}(1997)\citenamefont
  {Onodera}, \citenamefont {Amanai}, \citenamefont {Matsuo}, \citenamefont
  {Kosaka}, \citenamefont {Kobayashi}, \citenamefont {Ohashi},\ and\
  \citenamefont {Yamaguchi}}]{Onodera_1997}%
  \BibitemOpen
  \bibfield  {author} {\bibinfo {author} {\bibfnamefont {H.}~\bibnamefont
  {Onodera}}, \bibinfo {author} {\bibfnamefont {H.}~\bibnamefont {Amanai}},
  \bibinfo {author} {\bibfnamefont {S.}~\bibnamefont {Matsuo}}, \bibinfo
  {author} {\bibfnamefont {M.}~\bibnamefont {Kosaka}}, \bibinfo {author}
  {\bibfnamefont {H.}~\bibnamefont {Kobayashi}}, \bibinfo {author}
  {\bibfnamefont {M.}~\bibnamefont {Ohashi}}, \ and\ \bibinfo {author}
  {\bibfnamefont {Y.}~\bibnamefont {Yamaguchi}},\ }\href@noop {} {\bibfield
  {journal} {\bibinfo  {journal} {Scientific Reports of the Research
  Institutes, Tohoku University, Ser. A}\ }\textbf {\bibinfo {volume} {45}},\
  \bibinfo {pages} {1} (\bibinfo {year} {1997})}\BibitemShut {NoStop}%
\bibitem [{\citenamefont {Kolincio}\ \emph {et~al.}(2016)\citenamefont
  {Kolincio}, \citenamefont {P\'erez}, \citenamefont {H\'ebert}, \citenamefont
  {Fertey},\ and\ \citenamefont {Pautrat}}]{kolincio}%
  \BibitemOpen
  \bibfield  {author} {\bibinfo {author} {\bibfnamefont {K.}~\bibnamefont
  {Kolincio}}, \bibinfo {author} {\bibfnamefont {O.}~\bibnamefont {P\'erez}},
  \bibinfo {author} {\bibfnamefont {S.}~\bibnamefont {H\'ebert}}, \bibinfo
  {author} {\bibfnamefont {P.}~\bibnamefont {Fertey}}, \ and\ \bibinfo {author}
  {\bibfnamefont {A.}~\bibnamefont {Pautrat}},\ }\href {\doibase
  10.1103/PhysRevB.93.235126} {\bibfield  {journal} {\bibinfo  {journal} {Phys.
  Rev. B}\ }\textbf {\bibinfo {volume} {93}},\ \bibinfo {pages} {235126}
  (\bibinfo {year} {2016})}\BibitemShut {NoStop}%
\bibitem [{\citenamefont {Prathiba}\ \emph {et~al.}(2016)\citenamefont
  {Prathiba}, \citenamefont {Kim}, \citenamefont {Shin}, \citenamefont
  {Strychalska}, \citenamefont {Klimczuk},\ and\ \citenamefont
  {Park}}]{prathiba_tuning_2016}%
  \BibitemOpen
  \bibfield  {author} {\bibinfo {author} {\bibfnamefont {G.}~\bibnamefont
  {Prathiba}}, \bibinfo {author} {\bibfnamefont {I.}~\bibnamefont {Kim}},
  \bibinfo {author} {\bibfnamefont {S.}~\bibnamefont {Shin}}, \bibinfo {author}
  {\bibfnamefont {J.}~\bibnamefont {Strychalska}}, \bibinfo {author}
  {\bibfnamefont {T.}~\bibnamefont {Klimczuk}}, \ and\ \bibinfo {author}
  {\bibfnamefont {T.}~\bibnamefont {Park}},\ }\href {\doibase
  10.1038/srep26530} {\bibfield  {journal} {\bibinfo  {journal} {Scientific
  Reports}\ }\textbf {\bibinfo {volume} {6}},\ \bibinfo {pages} {26530}
  (\bibinfo {year} {2016})}\BibitemShut {NoStop}%
\bibitem [{\citenamefont {Shimomura}\ \emph {et~al.}(2009)\citenamefont
  {Shimomura}, \citenamefont {Hayashi}, \citenamefont {Asaka}, \citenamefont
  {Wakabayashi}, \citenamefont {Mizumaki},\ and\ \citenamefont
  {Onodera}}]{Shimomura_2009}%
  \BibitemOpen
  \bibfield  {author} {\bibinfo {author} {\bibfnamefont {S.}~\bibnamefont
  {Shimomura}}, \bibinfo {author} {\bibfnamefont {C.}~\bibnamefont {Hayashi}},
  \bibinfo {author} {\bibfnamefont {G.}~\bibnamefont {Asaka}}, \bibinfo
  {author} {\bibfnamefont {N.}~\bibnamefont {Wakabayashi}}, \bibinfo {author}
  {\bibfnamefont {M.}~\bibnamefont {Mizumaki}}, \ and\ \bibinfo {author}
  {\bibfnamefont {H.}~\bibnamefont {Onodera}},\ }\href {\doibase
  10.1103/PhysRevLett.102.076404} {\bibfield  {journal} {\bibinfo  {journal}
  {Phys. Rev. Lett.}\ }\textbf {\bibinfo {volume} {102}},\ \bibinfo {pages}
  {076404} (\bibinfo {year} {2009})}\BibitemShut {NoStop}%
\bibitem [{\citenamefont {Dieterich}\ and\ \citenamefont
  {Fulde}(1973)}]{Dieterich1973}%
  \BibitemOpen
  \bibfield  {author} {\bibinfo {author} {\bibfnamefont {W.}~\bibnamefont
  {Dieterich}}\ and\ \bibinfo {author} {\bibfnamefont {P.}~\bibnamefont
  {Fulde}},\ }\href {\doibase 10.1007/BF01397717} {\bibfield  {journal}
  {\bibinfo  {journal} {Zeitschrift f{\"u}r Physik A Hadrons and nuclei}\
  }\textbf {\bibinfo {volume} {265}},\ \bibinfo {pages} {239} (\bibinfo {year}
  {1973})}\BibitemShut {NoStop}%
\bibitem [{\citenamefont {Xu}\ \emph {et~al.}(2009)\citenamefont {Xu},
  \citenamefont {Bangura}, \citenamefont {Analytis}, \citenamefont {Fletcher},
  \citenamefont {French}, \citenamefont {Shannon}, \citenamefont {He},
  \citenamefont {Zhang}, \citenamefont {Mandrus}, \citenamefont {Jin},\ and\
  \citenamefont {Hussey}}]{xu_directional_2009}%
  \BibitemOpen
  \bibfield  {author} {\bibinfo {author} {\bibfnamefont {X.}~\bibnamefont
  {Xu}}, \bibinfo {author} {\bibfnamefont {A.~F.}\ \bibnamefont {Bangura}},
  \bibinfo {author} {\bibfnamefont {J.~G.}\ \bibnamefont {Analytis}}, \bibinfo
  {author} {\bibfnamefont {J.~D.}\ \bibnamefont {Fletcher}}, \bibinfo {author}
  {\bibfnamefont {M.~M.~J.}\ \bibnamefont {French}}, \bibinfo {author}
  {\bibfnamefont {N.}~\bibnamefont {Shannon}}, \bibinfo {author} {\bibfnamefont
  {J.}~\bibnamefont {He}}, \bibinfo {author} {\bibfnamefont {S.}~\bibnamefont
  {Zhang}}, \bibinfo {author} {\bibfnamefont {D.}~\bibnamefont {Mandrus}},
  \bibinfo {author} {\bibfnamefont {R.}~\bibnamefont {Jin}}, \ and\ \bibinfo
  {author} {\bibfnamefont {N.~E.}\ \bibnamefont {Hussey}},\ }\href {\doibase
  10.1103/PhysRevLett.102.206602} {\bibfield  {journal} {\bibinfo  {journal}
  {Physical Review Letters}\ }\textbf {\bibinfo {volume} {102}},\ \bibinfo
  {pages} {206602} (\bibinfo {year} {2009})}\BibitemShut {NoStop}%
\bibitem [{\citenamefont {Brooks}(2008)}]{brooks_magnetic_2008}%
  \BibitemOpen
  \bibfield  {author} {\bibinfo {author} {\bibfnamefont {J.~S.}\ \bibnamefont
  {Brooks}},\ }\href {\doibase 10.1088/0034-4885/71/12/126501} {\bibfield
  {journal} {\bibinfo  {journal} {Reports on Progress in Physics}\ }\textbf
  {\bibinfo {volume} {71}},\ \bibinfo {pages} {126501} (\bibinfo {year}
  {2008})}\BibitemShut {NoStop}%
\bibitem [{\citenamefont {Graf}\ \emph
  {et~al.}(2004{\natexlab{b}})\citenamefont {Graf}, \citenamefont {Brooks},
  \citenamefont {Choi}, \citenamefont {Uji}, \citenamefont {Dias},
  \citenamefont {Almeida},\ and\ \citenamefont
  {Matos}}]{graf_suppression_2004}%
  \BibitemOpen
  \bibfield  {author} {\bibinfo {author} {\bibfnamefont {D.}~\bibnamefont
  {Graf}}, \bibinfo {author} {\bibfnamefont {J.~S.}\ \bibnamefont {Brooks}},
  \bibinfo {author} {\bibfnamefont {E.~S.}\ \bibnamefont {Choi}}, \bibinfo
  {author} {\bibfnamefont {S.}~\bibnamefont {Uji}}, \bibinfo {author}
  {\bibfnamefont {J.~C.}\ \bibnamefont {Dias}}, \bibinfo {author}
  {\bibfnamefont {M.}~\bibnamefont {Almeida}}, \ and\ \bibinfo {author}
  {\bibfnamefont {M.}~\bibnamefont {Matos}},\ }\href {\doibase
  10.1103/PhysRevB.69.125113} {\bibfield  {journal} {\bibinfo  {journal}
  {Physical Review B}\ }\textbf {\bibinfo {volume} {69}},\ \bibinfo {pages}
  {125113} (\bibinfo {year} {2004}{\natexlab{b}})}\BibitemShut {NoStop}%
\bibitem [{\citenamefont {Kane-Maguire}\ \emph {et~al.}(2005)\citenamefont
  {Kane-Maguire}, \citenamefont {Officer}, \citenamefont {Graf}, \citenamefont
  {Choi}, \citenamefont {Brooks}, \citenamefont {Dias}, \citenamefont
  {Henriques}, \citenamefont {Almeida}, \citenamefont {Matos},\ and\
  \citenamefont {Rickel}}]{Graf2005}%
  \BibitemOpen
  \bibfield  {author} {\bibinfo {author} {\bibfnamefont {L.}~\bibnamefont
  {Kane-Maguire}}, \bibinfo {author} {\bibfnamefont {D.}~\bibnamefont
  {Officer}}, \bibinfo {author} {\bibfnamefont {D.}~\bibnamefont {Graf}},
  \bibinfo {author} {\bibfnamefont {E.}~\bibnamefont {Choi}}, \bibinfo {author}
  {\bibfnamefont {J.}~\bibnamefont {Brooks}}, \bibinfo {author} {\bibfnamefont
  {J.}~\bibnamefont {Dias}}, \bibinfo {author} {\bibfnamefont {R.}~\bibnamefont
  {Henriques}}, \bibinfo {author} {\bibfnamefont {M.}~\bibnamefont {Almeida}},
  \bibinfo {author} {\bibfnamefont {M.}~\bibnamefont {Matos}}, \ and\ \bibinfo
  {author} {\bibfnamefont {D.}~\bibnamefont {Rickel}},\ }\href {\doibase
  http://dx.doi.org/10.1016/j.synthmet.2005.07.306} {\bibfield  {journal}
  {\bibinfo  {journal} {Synthetic Metals}\ }\textbf {\bibinfo {volume} {153}},\
  \bibinfo {pages} {361 } (\bibinfo {year} {2005})}\BibitemShut {NoStop}%
\bibitem [{\citenamefont {Monchi}\ \emph {et~al.}(1999)\citenamefont {Monchi},
  \citenamefont {Poirier}, \citenamefont {Bourbonnais}, \citenamefont {Matos},\
  and\ \citenamefont {Henriques}}]{monchi_international_1999}%
  \BibitemOpen
  \bibfield  {author} {\bibinfo {author} {\bibfnamefont {K.}~\bibnamefont
  {Monchi}}, \bibinfo {author} {\bibfnamefont {M.}~\bibnamefont {Poirier}},
  \bibinfo {author} {\bibfnamefont {C.}~\bibnamefont {Bourbonnais}}, \bibinfo
  {author} {\bibfnamefont {M.~J.}\ \bibnamefont {Matos}}, \ and\ \bibinfo
  {author} {\bibfnamefont {R.~T.}\ \bibnamefont {Henriques}},\ }\href {\doibase
  10.1016/S0379-6779(98)00319-1} {\bibfield  {journal} {\bibinfo  {journal}
  {Synthetic Metals}\ }\textbf {\bibinfo {volume} {103}},\ \bibinfo {pages}
  {2228} (\bibinfo {year} {1999})}\BibitemShut {NoStop}%
\bibitem [{\citenamefont {Matos}\ \emph {et~al.}(1996)\citenamefont {Matos},
  \citenamefont {Bonfait}, \citenamefont {Henriques},\ and\ \citenamefont
  {Almeida}}]{matos_modification_1996}%
  \BibitemOpen
  \bibfield  {author} {\bibinfo {author} {\bibfnamefont {M.}~\bibnamefont
  {Matos}}, \bibinfo {author} {\bibfnamefont {G.}~\bibnamefont {Bonfait}},
  \bibinfo {author} {\bibfnamefont {R.~T.}\ \bibnamefont {Henriques}}, \ and\
  \bibinfo {author} {\bibfnamefont {M.}~\bibnamefont {Almeida}},\ }\href
  {\doibase 10.1103/PhysRevB.54.15307} {\bibfield  {journal} {\bibinfo
  {journal} {Physical Review B}\ }\textbf {\bibinfo {volume} {54}},\ \bibinfo
  {pages} {15307} (\bibinfo {year} {1996})}\BibitemShut {NoStop}%
\bibitem [{\citenamefont {Kittel}(2004)}]{kittel_2004}%
  \BibitemOpen
  \bibfield  {author} {\bibinfo {author} {\bibfnamefont {C.}~\bibnamefont
  {Kittel}},\ }\href@noop {} {\emph {\bibinfo {title} {Introduction to Solid
  State Physics}}},\ \bibinfo {edition} {8th}\ ed.\ (\bibinfo  {publisher}
  {Wiley},\ \bibinfo {address} {Hoboken},\ \bibinfo {year} {2004})\BibitemShut
  {NoStop}%
\bibitem [{\citenamefont {Kalvius}\ and\ \citenamefont
  {Kienle}(2012)}]{Nowik_2012}%
  \BibitemOpen
  \bibinfo {editor} {\bibfnamefont {M.}~\bibnamefont {Kalvius}}\ and\ \bibinfo
  {editor} {\bibfnamefont {P.}~\bibnamefont {Kienle}},\ eds.,\ \enquote
  {\bibinfo {title} {The {Rudolf} {Mossbauer} story: His scientific work and
  its impact on science and history},}\ \ (\bibinfo  {publisher} {Springer},\
  \bibinfo {year} {2012})\ Chap.\ \bibinfo {chapter} {The Internal Magnetic
  Fields Acting on Nuclei in Solids}\BibitemShut {NoStop}%
\bibitem [{\citenamefont {Bonfait}\ \emph {et~al.}(1995)\citenamefont
  {Bonfait}, \citenamefont {Matos}, \citenamefont {Henriques},\ and\
  \citenamefont {Almeida}}]{bonfait_research_1995-1}%
  \BibitemOpen
  \bibfield  {author} {\bibinfo {author} {\bibfnamefont {G.}~\bibnamefont
  {Bonfait}}, \bibinfo {author} {\bibfnamefont {M.~J.}\ \bibnamefont {Matos}},
  \bibinfo {author} {\bibfnamefont {R.~T.}\ \bibnamefont {Henriques}}, \ and\
  \bibinfo {author} {\bibfnamefont {M.}~\bibnamefont {Almeida}},\ }\href
  {\doibase 10.1016/0921-4526(94)01045-3} {\bibfield  {journal} {\bibinfo
  {journal} {Physica B: Condensed Matter}\ }\textbf {\bibinfo {volume} {211}},\
  \bibinfo {pages} {297} (\bibinfo {year} {1995})}\BibitemShut {NoStop}%
\bibitem [{\citenamefont {Kim}\ \emph {et~al.}(2012)\citenamefont {Kim},
  \citenamefont {Rhyee},\ and\ \citenamefont {Kwon}}]{Kim_2012}%
  \BibitemOpen
  \bibfield  {author} {\bibinfo {author} {\bibfnamefont {J.~H.}\ \bibnamefont
  {Kim}}, \bibinfo {author} {\bibfnamefont {J.-S.}\ \bibnamefont {Rhyee}}, \
  and\ \bibinfo {author} {\bibfnamefont {Y.~S.}\ \bibnamefont {Kwon}},\ }\href
  {\doibase 10.1103/PhysRevB.86.235101} {\bibfield  {journal} {\bibinfo
  {journal} {Phys. Rev. B}\ }\textbf {\bibinfo {volume} {86}},\ \bibinfo
  {pages} {235101} (\bibinfo {year} {2012})}\BibitemShut {NoStop}%
\bibitem [{\citenamefont {Karplus}\ and\ \citenamefont
  {Luttinger}(1954)}]{karplus_hall_1954}%
  \BibitemOpen
  \bibfield  {author} {\bibinfo {author} {\bibfnamefont {R.}~\bibnamefont
  {Karplus}}\ and\ \bibinfo {author} {\bibfnamefont {J.~M.}\ \bibnamefont
  {Luttinger}},\ }\href {\doibase 10.1103/PhysRev.95.1154} {\bibfield
  {journal} {\bibinfo  {journal} {Physical Review}\ }\textbf {\bibinfo {volume}
  {95}},\ \bibinfo {pages} {1154} (\bibinfo {year} {1954})}\BibitemShut
  {NoStop}%
\bibitem [{\citenamefont {Berger}\ and\ \citenamefont
  {Bergmann}(1980)}]{Berger1980}%
  \BibitemOpen
  \bibfield  {author} {\bibinfo {author} {\bibfnamefont {L.}~\bibnamefont
  {Berger}}\ and\ \bibinfo {author} {\bibfnamefont {G.}~\bibnamefont
  {Bergmann}},\ }\enquote {\bibinfo {title} {The hall effect of
  ferromagnets},}\ in\ \href {\doibase 10.1007/978-1-4757-1367-1_2} {\emph
  {\bibinfo {booktitle} {The Hall Effect and Its Applications}}},\ \bibinfo
  {editor} {edited by\ \bibinfo {editor} {\bibfnamefont {C.~L.}\ \bibnamefont
  {Chien}}\ and\ \bibinfo {editor} {\bibfnamefont {C.~R.}\ \bibnamefont
  {Westgate}}}\ (\bibinfo  {publisher} {Springer US},\ \bibinfo {address}
  {Boston, MA},\ \bibinfo {year} {1980})\ pp.\ \bibinfo {pages}
  {55--76}\BibitemShut {NoStop}%
\bibitem [{\citenamefont {Shiomi}\ \emph {et~al.}(2009)\citenamefont {Shiomi},
  \citenamefont {Onose},\ and\ \citenamefont {Tokura}}]{shiomi_extrinsic_2009}%
  \BibitemOpen
  \bibfield  {author} {\bibinfo {author} {\bibfnamefont {Y.}~\bibnamefont
  {Shiomi}}, \bibinfo {author} {\bibfnamefont {Y.}~\bibnamefont {Onose}}, \
  and\ \bibinfo {author} {\bibfnamefont {Y.}~\bibnamefont {Tokura}},\ }\href
  {\doibase 10.1103/PhysRevB.79.100404} {\bibfield  {journal} {\bibinfo
  {journal} {Physical Review B}\ }\textbf {\bibinfo {volume} {79}},\ \bibinfo
  {pages} {100404} (\bibinfo {year} {2009})}\BibitemShut {NoStop}%
\bibitem [{\citenamefont {Granovskii}\ \emph {et~al.}(2012)\citenamefont
  {Granovskii}, \citenamefont {Prudnikov}, \citenamefont {Kazakov},
  \citenamefont {Zhukov},\ and\ \citenamefont {Dubenko}}]{Granovskii2012}%
  \BibitemOpen
  \bibfield  {author} {\bibinfo {author} {\bibfnamefont {A.~B.}\ \bibnamefont
  {Granovskii}}, \bibinfo {author} {\bibfnamefont {V.~N.}\ \bibnamefont
  {Prudnikov}}, \bibinfo {author} {\bibfnamefont {A.~P.}\ \bibnamefont
  {Kazakov}}, \bibinfo {author} {\bibfnamefont {A.~P.}\ \bibnamefont {Zhukov}},
  \ and\ \bibinfo {author} {\bibfnamefont {I.~S.}\ \bibnamefont {Dubenko}},\
  }\href {\doibase 10.1134/S1063776112090051} {\bibfield  {journal} {\bibinfo
  {journal} {Journal of Experimental and Theoretical Physics}\ }\textbf
  {\bibinfo {volume} {115}},\ \bibinfo {pages} {805} (\bibinfo {year}
  {2012})}\BibitemShut {NoStop}%
\end{thebibliography}
%

\end{document}